\documentclass[12pt,aps,prd,preprint,tightenlines,showpacs,nofootinbib]{revtex4-1}
\newcommand{\PRE}[1]{{#1}} 

\usepackage{hyperref}      
\usepackage{bm} 
\usepackage{amsmath} 
\usepackage{epsfig}
\usepackage{subfigure}
\usepackage{dsfont}

\newcommand{\sigmaan}{\sigma_{\text{an}}}

\newcommand{\OmegaDM}{\Omega_{\text{DM}}}

\newcommand{\mev}{\text{MeV}}
\newcommand{\gev}{\text{GeV}}
\newcommand{\tev}{\text{TeV}}
\newcommand{\zb}{\text{zb}}
\newcommand{\pb}{\text{pb}}

\newcommand{\cm}{\text{cm}}

\newcommand{\eg}{{\em e.g.}}

\renewcommand{\eqref}[1]{Eq.~(\ref{#1})}
\newcommand{\eqsref}[2]{Eqs.~(\ref{#1}) and (\ref{#2})}
\newcommand{\Eqref}[1]{Equation~(\ref{#1})}
\newcommand{\secref}[1]{Sec.~\ref{sec:#1}}
\newcommand{\secsref}[2]{Secs.~\ref{sec:#1} and \ref{sec:#2}}

\newcommand{\figref}[1]{Fig.~\ref{fig:#1}}

\newcommand{\Figref}[1]{Figure~\ref{fig:#1}}

\newcommand{\mchi}{m_{\chi}}

\newcommand{\Bino}{\tilde{B}}
\newcommand{\Wino}{\tilde{W}}
\newcommand{\Higgsino}{\tilde{H}}

\newcommand{\slepton}{\tilde{\ell}}

\newcommand{\signmu}{\text{sign}(\mu)}
\newcommand{\Omegachi}{\Omega_{\Neut}}

\newcommand{\Neut}{\chi}
\newcommand{\mneut}{m_{\Neut}}
\newcommand{\mtilde}{\tilde{m}}
\newcommand{\sigmaSI}{\sigma^{\text{SI}}}
\newcommand{\sigmaSD}{\sigma^{\text{SD}}}
\newcommand{\ah}{a_{\Higgsino}}
\newcommand{\ahu}{a_{\Higgsino_u}}
\newcommand{\ahd}{a_{\Higgsino_d}}
\newcommand{\ahud}{a_{\Higgsino_{u,d}}}
\newcommand{\ab}{a_{\Bino}}
\newcommand{\aw}{a_{\Wino}}
\newcommand{\phiCP}{\phi_{\text{CP}}}
\newcommand{\disabled}[1]{{}}

\begin{document}

\preprint{UCI-TR-2010-16}

\title{ \PRE{\vspace*{1.5in}} 
Heart of Darkness: The Significance of the Zeptobarn Scale for
Neutralino Direct Detection \PRE{\vspace*{0.3in}}}

\author{Jonathan L.~Feng}
\affiliation{Department of Physics and Astronomy, University of
California, Irvine, California 92697, USA \PRE{\vspace*{.5in}}}

\author{David Sanford}
\affiliation{Department of Physics and Astronomy, University of
California, Irvine, California 92697, USA \PRE{\vspace*{.5in}}}

\date{September 2010}

\begin{abstract}
\PRE{\vspace*{.3in}} The direct detection of dark matter through its
elastic scattering off nucleons is among the most promising methods
for establishing the particle identity of dark matter. The current
bound on the spin-independent scattering cross section is $\sigmaSI <
10~\zb$ for dark matter masses $\mneut \sim 100~\gev$, with improved
sensitivities expected soon.  We examine the implications of this
progress for neutralino dark matter.  We work in a supersymmetric
framework well-suited to dark matter studies that is simple and
transparent, with models defined in terms of four weak-scale
parameters.  We first show that robust constraints on electric dipole
moments motivate large sfermion masses $\mtilde \agt 1~\tev$,
effectively decoupling squarks and sleptons from neutralino dark
matter phenomenology.  In this case, we find characteristic cross
sections in the narrow range $1~\zb \alt \sigmaSI \alt 40~\zb$ for
$\mneut \agt 70~\gev$.  As sfermion masses are lowered to near their
experimental limit $\mtilde \sim 400~\gev$, the upper and lower limits
of this range are extended, but only by factors of around two, and the
lower limit is not significantly altered by relaxing many particle
physics assumptions, varying the strange quark content of the nucleon,
including the effects of galactic small-scale structure, or assuming
other components of dark matter.  Experiments are therefore rapidly
entering the heart of dark matter-favored supersymmetry parameter
space.  If no signal is seen, supersymmetric models must contain some
level of fine-tuning, and we identify and analyze several
possibilities.  Barring large cancellations, however, in a large and
generic class of models, if thermal relic neutralinos are a
significant component of dark matter, experiments will discover them
as they probe down to the zeptobarn scale.
\end{abstract}


\maketitle

\section{Introduction}
\label{sec:intro}

Now is an exciting time for the study of non-baryonic cold dark matter
(DM).  The existence of DM offers a consistent explanation for a wide
variety of astrophysical observations on disparate length scales.  At
the same time, there are many viable dark matter candidates,
motivating a diverse and active research program searching for the
direct detection of dark matter scattering off nuclei, the indirect
detection of dark matter annihilation or decay products, dark matter
production at colliders, and the effects of dark matter particle
properties on structure formation and other astrophysical
phenomena~\cite{Bertone:2004pz,Bergstrom:2009ib,Feng:2010gw}.

In this work, we focus on direct detection.  We will discuss and
present results for both spin-independent and spin-dependent
scattering, but will focus particularly on probes of the
spin-independent dark matter-nucleon scattering cross section
$\sigmaSI$.  Although no definitive signal of dark matter has been
established so far, recent results from the
DAMA~\cite{Bernabei:2010mq}, XENON10~\cite{Angle:2008we},
CDMS-II~\cite{Ahmed:2009zw}, CoGeNT~\cite{Aalseth:2010vx}, and
XENON100~\cite{Aprile:2010um, Aprile:2011hx, Aprile:2011hi}
experiments have sparked great interest.  Moreover, significant
progress for spin-independent cross sections is expected from several
sources, including XENON100~\cite{Aprile:2009yh},
CRESST~\cite{Schmaler:2009bu}, LUX~\cite{McKinsey:2010zz},
XMASS~\cite{Sekiya:2010bf}, mini-CLEAN/DEAP~\cite{Hime:2006zq},
WArP~\cite{Acciarri:2010zz}, and superCDMS~\cite{Bruch:2010eq}.
Current limits are so stringent that there is no agreed upon unit
convention for expressing them.  We advocate the use of zeptobarns
($1~\zb = 10^{-9}~\pb = 10^{-45}~\cm^2$), which is more compact and no
less transparent than the alternatives. The current limit is $\sigmaSI
< 10~\zb$ for 100 GeV dark matter~\cite{Ahmed:2009zw,Aprile:2010um,
  Aprile:2011hi}.  In the near future, results from 100 kg-scale
detectors may improve this sensitivity by an order of magnitude to the
zeptobarn scale.  Future ton-scale detectors are being planned to
extend sensitivities even beyond that, with the ultimate goal of
probing yoctobarn cross sections, where experiments will reach the
background-free limit for non-directional detectors imposed by
irreducible neutrino backgrounds~\cite{Strigari:2009bq,Gutlein:2010tq}
(and also the outer limits of the metric prefix system).

The single most studied dark matter candidate is the lightest
neutralino $\Neut$ predicted by supersymmetry
(SUSY)~\cite{Goldberg:1983nd,Ellis:1983ew,Drees:1992am,%
Drees:1993bu,Jungman:1995df}. SUSY provides a solution to the gauge
hierarchy problem and gauge coupling unification, and with $R$-parity
conservation, the lightest supersymmetric particle (LSP) is stable.
The LSP in SUSY theories is often the lightest neutralino, a mixture
of the superpartners of the neutral electroweak gauge and Higgs
bosons.  It is an excellent DM candidate, not only because SUSY is a
well-motivated possibility for new physics, but also because the
lightest neutralino is a weakly-interacting massive particle (WIMP),
and so naturally has the appropriate thermal relic density.

What are the possible values of $\sigmaSI$ for neutralino dark matter?
In full generality, the allowed range varies from very large values
that are far above current experimental
bounds~\cite{Bottino:2008mf,Feldman:2010ke,Kuflik:2010ah} to very
small values that are far below the sensitivity of foreseeable
experiments~\cite{Mandic:2000jz}.  At the same time, there is some
sense in which the extreme values occur only in special regions of
SUSY parameter space.  In this study, we aim to answer a more subtle
(and less well-defined) question: what are the {\em characteristic}
values of $\sigmaSI$ for neutralinos?  As experiments improve, is
every factor-of-ten improvement in $\sigmaSI$ equally important, or
are there certain thresholds that should be considered particularly
significant?  In other words, what is $\sigmaSI$ in the heart of SUSY
parameter space?

To answer these questions, one must define preferred models, reducing
the general parameter space to some subspace by appealing to some
organizational principle.  One approach is to work with a low-energy
effective theory that includes the dark matter particle but not other
supersymmetric degrees of freedom.  Dark matter-quark couplings are
included with, for example, four-fermion interactions $\bar{\Neut}
\Neut \bar{q} q$, suppressed by some large mass scale. This approach
has the advantage of being model-independent, with results applicable
to theories beyond SUSY, and has been adopted in a number of recent
studies motivated by the possibility of light (GeV-scale)
WIMPs~\cite{Goodman:2010yf,Goodman:2010ku,Bai:2010hh,Fan:2010gt}.  The
effective theory approach is, however, inapplicable to the generic
SUSY theories of interest to us, in which the dark matter mass is not
hierarchically smaller than those of the other superparticles.  Also,
effective theory studies typically assume that the same operators
govern both annihilation and scattering, and so neglect the
annihilation to leptons and gauge bosons that may be important in SUSY
theories.

Another approach has been to focus on a particular high-energy
framework, incorporating constraints from a battery of known
observables, and scanning over the reduced, but still high
dimensional, parameter space. The most popular framework for these
analyses is minimal supergravity (mSUGRA)~\cite{Chamseddine:1982jx,%
  Barbieri:1982eh,Ohta:1982wn,Hall:1983iz,AlvarezGaume:1983gj}, with
its cosmologically-preferred regions, including the co-annihilation
region at low universal scalar mass
$m_0$~\cite{Ellis:1999mm,Baer:2002fv}, the focus point region at high
$m_0$~\cite{Feng:2000gh,Feng:2000zu,Baer:2005ky}, and the $A$-funnel
region at high $\tan\beta$~\cite{Ellis:2001msa,Baer:2002fv}.  Much
work has been devoted to this framework; for recent analyses, see,
\eg,
Refs.~\cite{deAustri:2006pe,Trotta:2006ew,Trotta:2008bp,Baer:2006te}.
This approach has the advantage that there is a manifestly consistent
high-energy formulation throughout parameter space, but there are also
drawbacks.  The relation of the fundamental high-energy parameters to
physical observables, such as superpartner masses, is typically
opaque, at least to non-experts.  In addition, the framework allows
one to relate dark matter properties to $b \to s \gamma$, the
anomalous magnetic moment of the muon, and other {\em a priori}
unrelated observables, and it is not clear how robust these
connections should be.  Finally, not all of the assumptions of mSUGRA
are well-motivated, and it is not always easy to determine how dark
matter results would change if some of the assumptions were relaxed or
another framework adopted.  For example, are the promising predictions
for $\sigmaSI$ in focus point supersymmetry necessarily tied to large
scalar masses, with their negative implications for colliders?  Are
there consequences of the connection between sfermion and Higgs masses
in mSUGRA that fail to translate to more general theories?

In this work we take a complementary approach.  Although we will apply
some constraints and characteristics from high-energy motivations, we
work in a manifestly low-energy framework, with models defined by a
few weak-scale SUSY parameters that are easily related to physical
observables, typically the physical masses of certain superpartners.
The model framework is particularly well-suited to dark matter studies
for the reasons given below, and although specific to SUSY, our
results will apply to a broad class of SUSY models.  In addition, it
will allow us to explore the entire parameter space and understand the
physics behind the results in a detailed and relatively transparent
way.  This in some ways resembles the approach taken in the so-called
phenomenological MSSM (pMSSM) \cite{Djouadi:1998di, Djouadi:2002ze,
  Berger:2008cq, Cotta:2009zu}, though our exclusive interest in
neutralino dark matter allows for a more directed study of a smaller
parameter space.  We will find that many correlations found in mSUGRA
analyses are broken in these more phenomenological models.  In
particular, the large signals for direct detection found in focus
point supersymmetry are generic and not necessarily tied to large
scalar masses, with highly favorable implications for the immediate
future of neutralino direct detection.

This paper is organized as follows.  In \secref{model}, we define our
model framework and discuss its input parameters and assumptions.  We
also discuss the most relevant constraints, notably those from bounds
on the electric dipole moments (EDMs) of the electron and neutron.  In
\secref{relic} we analyze the dependence of the thermal relic density
on the model parameters, and in \secref{dd} we briefly review the form
of $\sigmaSI$ and explain its expected behavior in our model
framework.  In \secref{main} we present our main results for the
characteristic values of $\sigmaSI$, the possibility of fine-tuned
cancellations, and the implications for dark matter discovery.  In
\secref{caveats} we explore the effects of relaxing various particle
physics, nuclear physics, and astrophysical assumptions, and we
discuss the prospects for spin-dependent scattering in \secref{sd}.
We summarize our conclusions in \secref{conclusion}.

\section{Model Framework}
\label{sec:model}

\subsection{Input Parameters and Assumptions}
\label{sec:params}

We consider SUSY models with minimal field content that are specified
by five parameters defined at the weak scale $\sim 1~\tev$:
\begin{equation}
\label{eqn:params}
\mtilde, M_1, \mu, m_A, \tan\beta \ .
\end{equation}
In order, these are the unified sfermion soft SUSY-breaking mass, the
Bino mass, the Higgsino mass, the pseudoscalar Higgs boson mass, and
the ratio of Higgs boson expectation values.  

The rest of the model is specified by the following three assumptions:
\begin{enumerate}
\setlength{\itemsep}{0pt}\setlength{\parskip}{0pt}\setlength{\parsep}{0pt}
\item Gaugino mass unification.  We assume that the gaugino masses are
unified at the grand unified theory (GUT) scale, implying the
weak-scale gaugino mass relation
\begin{equation}
M_1 : M_2 : M_3 \simeq 1 : 2 : 6 \ .
\end{equation}

\item No left-right mixing.  We assume there is no significant
  left-right sfermion mixing, which fixes the ``$A$-term'' tri-linear
  scalar couplings.  For example, the top squark mass matrix
  simplifies as follows:
\begin{equation}
\left( \begin{array}{cc}
m_{\tilde{Q}}^2 + m_t^2 & m_t \left(A_t - \mu \cot \beta \right) \\
m_t \left(A_t - \mu \cot \beta \right) & m_{\tilde{u}}^2 + m_t^2
\end{array} \right)
\to
\left( \begin{array}{cc}
\mtilde^2 + m_t^2 & 0 \\ 
0 & \mtilde^2 + m_t^2
\end{array} \right) \ .
\end{equation}

\item Correct electroweak symmetry breaking.  This fixes the soft
Higgs scalar mass parameters $m_{H_{u,d}}^2$ and $m_{12}^2$ through
the (tree-level) conditions
\begin{eqnarray}
\mu^2 &=& \frac{m_{H_d}^2 - m_{H_u}^2 \tan^2 \beta + \frac{1}{2} m_Z^2
  \left(1 - \tan^2 \beta \right)}{\tan^2 \beta - 1} \\
m_A^2 &=& m_{H_d}^2 + m_{H_u}^2 + 2 \mu^2 \\
m_{12}^2 &=& m_A^2 \sin\beta \cos \beta \ .
\end{eqnarray}

\end{enumerate}

The model framework defined above is simple, but well-suited to
studies of neutralino dark matter.  With the assumption of gaugino
mass unification, the parameters $M_1$, $\mu$, and $\tan\beta$
completely determine the neutralino mass matrix, and so are natural
input parameters for this study.  These also fix the chargino mass
matrix and the gluino mass.  The remaining input parameters $\mtilde$
and $m_A$ determine the masses of all the remaining particles.  In our
numerical work, we use SOFTSUSY~3.1.2~\cite{Allanach:2001kg} to
generate the superpartner spectrum and
micrOMEGAS~2.2~\cite{Belanger:2006is,Belanger:2008sj} for relic
density and direct detection calculations.

We conclude this subsection with a few further comments about the
model parameter space and assumptions.  Throughout this analysis, we
will assume that the neutralino freezes out with the correct thermal
relic density $\Omegachi = \OmegaDM \simeq 0.23$.  Our results will be
remarkably insensitive to this assumption, as discussed in
\secref{astro}, but this constraint further simplifies the parameter
space.  As discussed in \secref{relic}, for a fixed $\mchi$, the relic
density determines $\mu$ up to a discrete choice, which may be taken
to be the sign of $\mu$.  The input parameters then become
\begin{equation}
\mtilde, M_1, \mu, m_A, \tan\beta \xrightarrow{\Omegachi = 0.23} 
\mchi ; \mtilde, \signmu, m_A, \tan\beta \ .
\end{equation}
We will typically plot quantities as functions of $\mchi$ for the two
choices of $\signmu$ and fixed $\mtilde$, $m_A$, and $\tan\beta$.

The assumption of gaugino mass unification is motivated by the
unification of gauge couplings and forces, a key virtue of
supersymmetry.  It implies that the LSP does not have a large Wino
component, which restricts the generality of this analysis.  On the
other hand, a neutralino with a significant Wino component typically
annihilates too efficiently to have the correct thermal relic density,
and so our assumption that the neutralino is a thermal relic already
disfavors this possibility~\cite{Mizuta:1992ja}.  Moreover, while
neutralino dark matter with non-unified gaugino masses has been
discussed~\cite{Baer:2006dz,Baer:2007xd}, and more complicated
GUT-breaking schemes are
possible~\cite{Anderson:1996bg,Chattopadhyay:2001va}, the gaugino mass
unification scenario used here remains a standard assumption for many
high-scale models.  However, it is worth noting that our results do
not require strict gaugino mass unification, but only that $M_2$ and
$M_3$ are at least somewhat larger than $M_1$.

The assumptions of a unified weak-scale sfermion mass $\mtilde$ and
negligible left-right mixing are not motivated by high-energy theories
of SUSY-breaking; instead, they are motivated by the relative
insensitivity of dark matter phenomenology to these assumptions.  We
will explore the implications of deviations from the above assumptions
in \secsref{splitmasses}{LR}, respectively.  We note here, however,
that the dependence on left-right mixing is small, and the neutralino
interactions of greatest interest here depend primarily on the squarks
in most regions of parameter space~\footnote{The relic density only
  depends strongly on slepton masses in the presence of slepton
  co-annihilation or very light slepton masses.}.  In many theories
the assumption of a unified squark mass is reasonably accurate, and so
this parametrization characterizes these models well, with $\mtilde$
playing the role of the characteristic squark mass.  Throughout this
study, we neglect sfermion flavor mixing and one-loop corrections to
sfermion masses, as these are sub-dominant effects in neutralino
interactions.

Finally, in our model framework, correct electroweak symmetry breaking
is achieved at the expense of unifying the soft Higgs scalar masses
with those of the squarks and sleptons.  The unification of the Higgs
scalar masses with the soft squark and slepton masses is assumed in
mSUGRA, but it is one of the least motivated aspects of that
framework.  Flavor constraints motivate flavor-independent squark and
slepton masses, but do not constrain the Higgs scalar masses.  Even in
GUTs, the Higgs scalars and sfermions are typically in different
multiplets.  Our assumption regarding the Higgs mass parameters
follows other analyses of non-unified Higgs
models (NUHM)~\cite{Ellis:2002wv,BirkedalHansen:2002am,Ellis:2003eg,%
  Baer:2005bu}.

\subsection{Constraints}

\subsubsection{Flavor and CP Violation}
\label{sec:cp}

Theories that introduce new particles at the weak scale run the risk
of violating stringent bounds on low-energy flavor- and CP-violating
observables.  For SUSY, there are mediation mechanisms that lead to
generation-independent slepton and squark masses, which allow these
theories to satisfy constraints on flavor violation.  These will also
be satisfied in our models, given the assumption of a unified
weak-scale sfermion mass $\mtilde$.

Even in these cases, however, constraints from EDMs, which conserve
flavor but violate CP, are not necessarily satisfied.  In a general
SUSY model, the EDMs of the electron and neutron receive contributions
from several sources.  Left-right sfermion mixing can lead to CP
violation, but this is absent in our model framework.  However, there
are also contributions from penguin diagrams with gauginos, Higgsinos,
and a sfermion of fixed chirality in the loop, which are proportional
to $\sin \phiCP$, where $\phiCP$ is the mismatch in phases of the
gaugino masses and $\mu$ parameter.  In the gauge eigenstate basis,
the dominant diagrams are those with neutral Winos, neutral Higgsinos,
and left-handed sfermions in the loop, and a similar one with charged
Winos, charged Higgsinos, and sneutrinos in the loop.  The first
diagram leads to the EDM contribution~\cite{Feng:2001sq}
\begin{equation}
d_f = \frac{1}{4} e \, m_f \, g_2^2 \, |M_2 \mu| \, \tan \beta \, 
\sin \phiCP \, 
K_N ( m_{\tilde{f}_L}^2, |\mu|^2, |M_2|^2) \ ,
\end{equation}
where $K_N$ is a kinematic function that may be approximated as $K_N
\sim 1/(16 \pi^2 m_{\tilde{f}_L}^4)$~\cite{Moroi:1995yh}. 

The current bounds on the electron and neutron EDMs are $d_e < 1.6
\times 10^{-27}~e~\cm$~\cite{Regan:2002ta} and $d_n < 2.9 \times
10^{-26}~e~\cm$~\cite{Baker:2006ts}.  Assuming $m_u = 3~\mev$, $m_d =
5~\mev$, the naive quark model relation $d_n = (4 d_d - d_u)/3$, and
neglecting cancellations between different diagrams, we find
\begin{eqnarray}
\frac{|M_2 \mu|}{m_{\tilde{e}_L}^2} 
\left[ \frac{6.5~\tev}{m_{\tilde{e}_L}} \right]^2
\frac{\tan \beta}{10} \sin \phiCP & \alt & 1 \\
\frac{|M_2 \mu|}{m_{\tilde{u}_L}^2}
\left[ \frac{2.1~\tev}{m_{\tilde{u}_L}} \right]^2
\frac{\tan \beta}{10} \sin \phiCP & \alt & 1 \\
\frac{|M_2 \mu|}{m_{\tilde{d}_L}^2}
\left[ \frac{5.5~\tev}{m_{\tilde{d}_L}} \right]^2
\frac{\tan \beta}{10} \sin \phiCP & \alt & 1 \ .
\end{eqnarray}
In our model context, with unified scalar masses and $M_2 \simeq 2
M_1$, these imply
\begin{equation}
\frac{|M_1 \mu|}{\mtilde^2}
\left[ \frac{9.2~\tev}{\mtilde} \right]^2
\frac{\tan \beta}{10} \sin \phiCP \alt 1 \ .
\end{equation}
Note, however, that even without assuming unified slepton and squark
masses, the constraints on $m_{\tilde{e}_L}$ and $m_{\tilde{d}_L}$ are
similar, and the constraint on $m_{\tilde{u}_L}$ is only somewhat less
restrictive.

This is a robust and numerically stringent constraint.  Even for the
low values $|M_1|, |\mu| \sim 100~\gev$ and moderate $\tan \beta \sim
10$, this implies $[ 1~\tev / \mtilde ]^4 \, \sin \phiCP \alt 1$.
Without a mechanism for aligning the CP-violating phases of different
SUSY parameters, we naturally expect $\sin \phiCP \sim 1$, implying
$\mtilde \agt 1~\tev$.  Lower values of $\mtilde$ are possible, of
course, but require small values of $\phiCP$.  As we will see, for
$\mtilde \agt 1~\tev$, the sfermions are largely decoupled from
neutralino interactions, leading to many simplifications in the
analysis of neutralino annihilation and scattering.  We will present
results for a variety of $\mtilde$ up to 2 TeV.  For this largest
value, arbitrary CP-violating phases are allowed for $|M_1|, |\mu|
\sim 400~\gev$, and only moderate suppression from $\sin \phiCP \alt
0.2$ is necessary for $|M_1|, |\mu| \alt 1~\tev$.

There are several other constraints on flavor and CP violation that
are often used to evaluate the viability of SUSY theories.  The most
notable are the constraints on the anomalous magnetic moment of the
muon $a_{\mu} = (g_{\mu}-2)/2$~\cite{Bennett:2006fi} and the rare
decays $b \to s \gamma$~\cite{Barberio:2007cr} and $b \to s \mu^+
\mu^-$~\cite{Aaltonen:2007kv}.  We will neglect these constraints in
our analysis, as large sfermion masses motivated by EDMs largely avoid
these constraints.  The experimental value for $B (b \to s \gamma)$ is
consistent with the SM prediction~\cite{Misiak:2006zs,Misiak:2006ab},
and $B (b \to s \mu^+ \mu^-)$ has only an upper limit.  In both cases,
large sfermion masses suppress the SUSY contributions to these
observables sufficiently to avoid the constraints.  The contribution
to $a_{\mu}$ is also suppressed for large sfermion masses, implying
that models in our framework do not alleviate the $a_{\mu}$
discrepancy~\cite{Jegerlehner:2009ry}, but do not exacerbate it
either.

\subsubsection{Higgs Boson Mass}
\label{sec:higgs}

Since our models are defined in terms of weak-scale parameters, their
implications for physical masses are typically clear, and we simply
avoid regions of parameter space that are excluded by direct searches
for new particles.  

The one exception to this rule is the lighter CP-even Higgs boson $h$.
At tree level, $m_h \approx m_Z$ for $\tan\beta \agt 10$ and $m_A \gg
m_Z$, and raising it above the LEP bound of $114.4~\gev$ requires
significant one-loop corrections involving top squarks.  In the
minimal supersymmetric standard model (MSSM), for $m_{\tilde{t}} \alt
1~\tev$, significant left-right top squark mixing is required to boost
$m_h$ to acceptable levels.  In non-minimal models, however, it is
possible to increase the lightest Higgs mass in other ways.  For
example, in the next-to-minimal supersymmetric standard model with a
singlet scalar mass of $M_S \sim 5 \mu$, the Higgs boson mass may be
as large as $m_h \approx 125~\gev$~\cite{Dine:2007xi}, even in cases
where the MSSM contribution leaves $m_h \sim m_Z$.  Although this
introduces a new complex scalar and a neutralino that in general mixes
with the rest of the neutralinos, the mass is sufficiently high that
it effectively decouples.

For these reasons, we set $m_h = 114.4~\gev$ whenever the explicit
one-loop corrections of our model are insufficient to drive it above
that bound.  Note that this redefinition is primarily important for
sfermion masses disfavored by the EDM constraints discussed above.  In
addition, raising the Higgs boson mass generally decreases $\sigmaSI$,
so simply removing models with too-low $m_h$ would, if anything, raise
the lower limit of our characteristic range of $\sigmaSI$.

\subsubsection{Extrapolation to High Scales}
\label{sec:extrapolation}

Because these models are specified by weak-scale input parameters,
there may be tensions associated with extrapolating them to high
energies through renormalization group evolution.  Studies of
non-unified Higgs models~\cite{Ellis:2002wv,Baer:2005bu} have found
that particular choices of $\mu$ and $m_A$ produce an unstable Higgs
potential at the high scale due to renormalization effects.  This
problem is generally avoided for $\mu \alt 1~\tev$ and large $m_A$.
The exact requirement for $m_A$ depends on other model parameters, but
$m_A \agt 1~\tev$ is generally large enough to avoid this difficulty.
A moderately large value of $m_A \agt 300 - 500~\gev$ is also
motivated by the desire for a sufficiently large relic density, with
low values generically producing overly-efficient annihilation even
without resonance effects.

Large gluino masses, with $M_3 \gg m_{\tilde{q}}$, may also imply
tachyonic squarks at the high scale.  This is not necessarily a
problem, as the low-energy theory could be located in a false vacuum
that is stable on cosmological time scales, a possibility that has
been considered in several studies~\cite{Falk:1995cq,Kusenko:1996jn,%
Feng:2005ba,Rajaraman:2006mr,Ellis:2008mc,Evans:2008zx}.  Of course,
all of these potential tensions assume a desert above the TeV scale
and may be avoided by the introduction of new heavy fields without
altering the phenomenological analysis presented here.

\section{Relic Density}
\label{sec:relic}

In this study, we require that the neutralino be a thermal relic, with
$\Omegachi = 0.23$.  In this section, we analyze what this constraint
implies for the gaugino and Higgsino content of the neutralino LSP,
which will help us understand the results for neutralino scattering
cross sections.

The mass eigenstate $\Neut$ is a mixture of gaugino and Higgsino
eigenstates given by
\begin{equation}
\Neut = \ab \Bino + \aw \Wino + \ahu \Higgsino_u + \ahd \Higgsino_d \ .
\end{equation}
With the assumption of gaugino mass unification, $\aw\ll 1$, since
$M_2 \simeq 2 M_1$, and so $\Neut$ is primarily a Bino-Higgsino
mixture.  The exact forms for $\ahud$ are~\cite{ElKheishen:1992yv}
\begin{eqnarray}
\frac{\ahd}{\ab} & = & \frac{\mu \left(M_2 \! -
  \! \mneut \right) \left(M_1 \! - \! \mneut \right) - \frac{1}{2} m_Z^2
  \sin 2\beta \left[ \left( M_1 \! - \! M_2 \right) \cos^2 \theta_W + M_2 -
  \mneut \right]}{m_Z \left(M_2 - \mneut \right) \sin\theta_W
  \left(\mu \cos\beta + \mneut \sin\beta\right)} \\
\frac{\ahu}{\ab} & = & \frac{- \mneut \left(M_2 \! -
  \! \mneut \right) \left(M_1 \! - \! \mneut \right) - m_Z^2 \cos^2 \beta
  \left[ \left( M_1 \! - \! M_2 \right) \cos^2 \theta_W + M_2 - \mneut
  \right]}{m_Z \left(M_2 - \mneut \right) \sin\theta_W \left(\mu
  \cos\beta + \mneut \sin\beta\right)} \ .
\end{eqnarray}
For moderate to large $\tan\beta$, both $\sin 2\beta$ and $\cos^2
\beta$ are small, and the first terms in both numerators are the
dominant contributions.  This conclusion may be evaded only if $\mu
\ll m_Z$, which predicts a light Higgsino and is excluded by LEP
bounds, or if $M_1$ and $\mneut$ are almost exactly degenerate, in
which case $\Neut \approx \Bino$ and $|\ahud| \ll 1$.  Thus, either
$\Neut \approx \Bino$ and $|\ahud| \ll 1$, or $\Neut$ is a significant
mixture of gaugino and Higgsino, and
\begin{equation}
\label{eqn:ahud}
\frac{\ahd}{\ahu} \approx - \frac{\mu}{\mneut} \sim {\cal O}(1) \ .
\end{equation}
Given this fact, it is useful to define a single parameter, the
``Higgsino-ness''
\begin{equation}
\ah \equiv \sqrt{\ahu^2 + \ahd^2} \ ,
\end{equation}
which characterizes the size of both $\ahu$ and $\ahd$ when they are
significant.  Note that there is also an important dependence on the
sign of $\mu$, which determines the relative signs of $\ab$ and
$\ahud$.  We will generally find solutions to $\Omegachi = 0.23$ for
both $\mu > 0$ and $\mu < 0$, and so we will present results for both
sign choices.

Discussion of $\Bino$-$\Higgsino$ mixing is familiar from focus point
supersymmetry~\cite{Feng:2000gh,Baer:2005ky} and well-tempered
neutralino scenarios~\cite{ArkaniHamed:2006mb,Baer:2006te}.  However,
we wish to stress that this mixing can be regarded as the key
parameter for all choices of parameters~\footnote{Even models without
  a unified gaugino mass, the $\Wino$ component of $\Neut$ is small so
  long as $M_2$ is even moderately larger than $M_1$.}, even in the
other typically-defined ``bulk'', ``co-annihilation'', and
``A-funnel'' dark matter regions discussed in mSUGRA.  Our approach,
however, effectively separates this mixing from the correspondence of
$\mu$ with scalar masses found in mSUGRA~\footnote{This is a general
  feature of NUHM scenarios~\cite{Ellis:2002wv,Baer:2005bu}}, as well
as avoiding other effects of running based on high-scale parameters
which obfuscate the electroweak scale mass and mixing parameters.

\begin{figure}[tbp]
\begin{center}
 \unitlength = 0.75mm
 \subfigure[Chargino-mediated Annihilation to $W^+ W^-$]{
   \includegraphics[width=.3\textwidth]{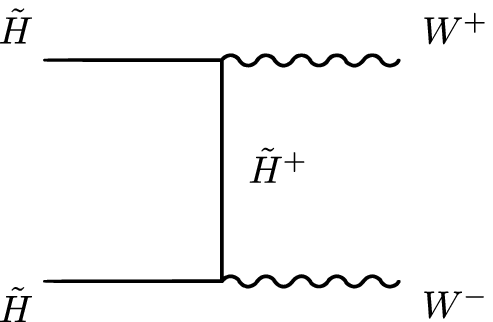}
   \label{fig:annihilationWW} }
 \subfigure[Higgs-mediated Annihilation to $W^+ W^-$]{
   \includegraphics[width=.3\textwidth]{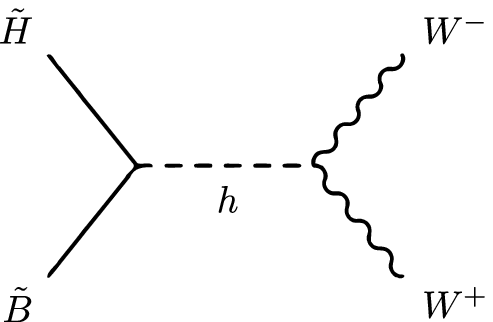}
   \label{fig:annihilationWWHiggs} }
 \subfigure[Higgs-mediated Annihilation to Fermions]{
   \includegraphics[width=.3\textwidth]{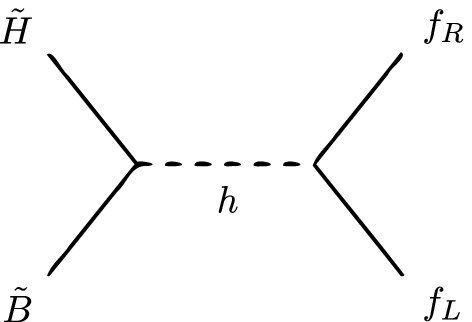}
   \label{fig:annihilationQQHiggs} }
 \subfigure[Sfermion-mediated Annihilation to Fermions]{
   \includegraphics[width=.3\textwidth]{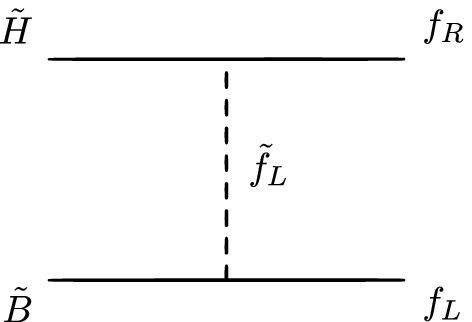}
   \label{fig:annihilationQQ} }
 \subfigure[Sfermion-mediated Annihilation to Fermions with LR Mixing]{
   \includegraphics[width=.3\textwidth]{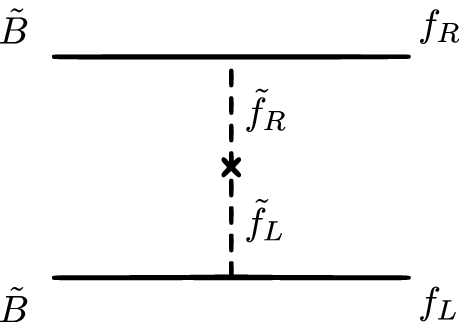}
   \label{fig:annihilationQQLR} }
\end{center}
\caption{{\em Major processes contributing to S-wave $\Neut \Neut$
    annihilation.}  The diagrams are labeled with the appropriate
  gauge eigenstates of the neutralino to highlight the required
  $\Bino$ and $\Higgsino$ contributions.  In
  \subref{fig:annihilationQQ} and \subref{fig:annihilationQQLR},
  additional diagrams exist with the exchange $\Bino \leftrightarrow
  \Higgsino$ and either $f_L \leftrightarrow f_R$ or $\tilde{f}_L
  \leftrightarrow \tilde{f}_R$.}
\label{fig:annihilation}
\end{figure}

The dominant processes contributing to $\Neut \Neut$ annihilation in
most regions of parameter space are shown in
\figref{annihilation}.\footnote{Generally other $t$-channel processes,
such as annihilation to $ZZ$, are sub-dominant.  There are, however,
specific regions of parameter space where resonant annihilation
dominates for an intermediate particle of mass $m_I \approx 2 \mneut$
or where co-annihilation is a major effect. We include all of these
effects in our numerical results.} The annihilation processes shown in
\figref{annihilationWW}-\figref{annihilationQQ} require a $\Neut$ with
a $\Higgsino$ component and thus scale with $\ah$.  The remaining
process of \figref{annihilationQQLR} does not require a $\Higgsino$
component, but it is generally suppressed relative to
\figref{annihilationQQ} because it requires left-right sfermion
mixing.

\begin{figure}[tbp]
  \subfigure[$\mtilde = 2~\tev$]{
    \includegraphics[width=.48\textwidth]{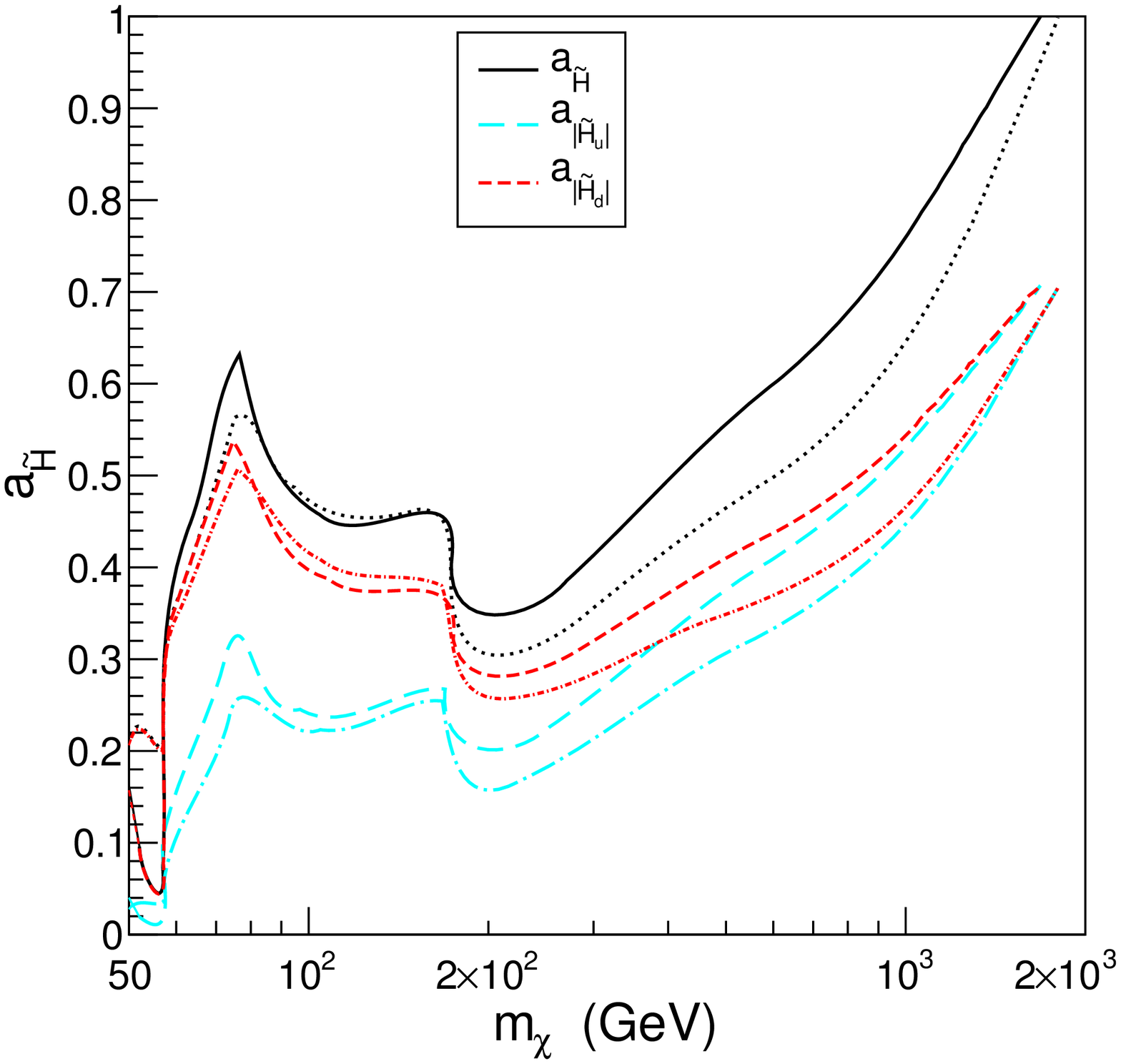}
    \label{fig:ah2TeV} }
  \subfigure[Various $\mtilde$]{
    \includegraphics[width=.48\textwidth]{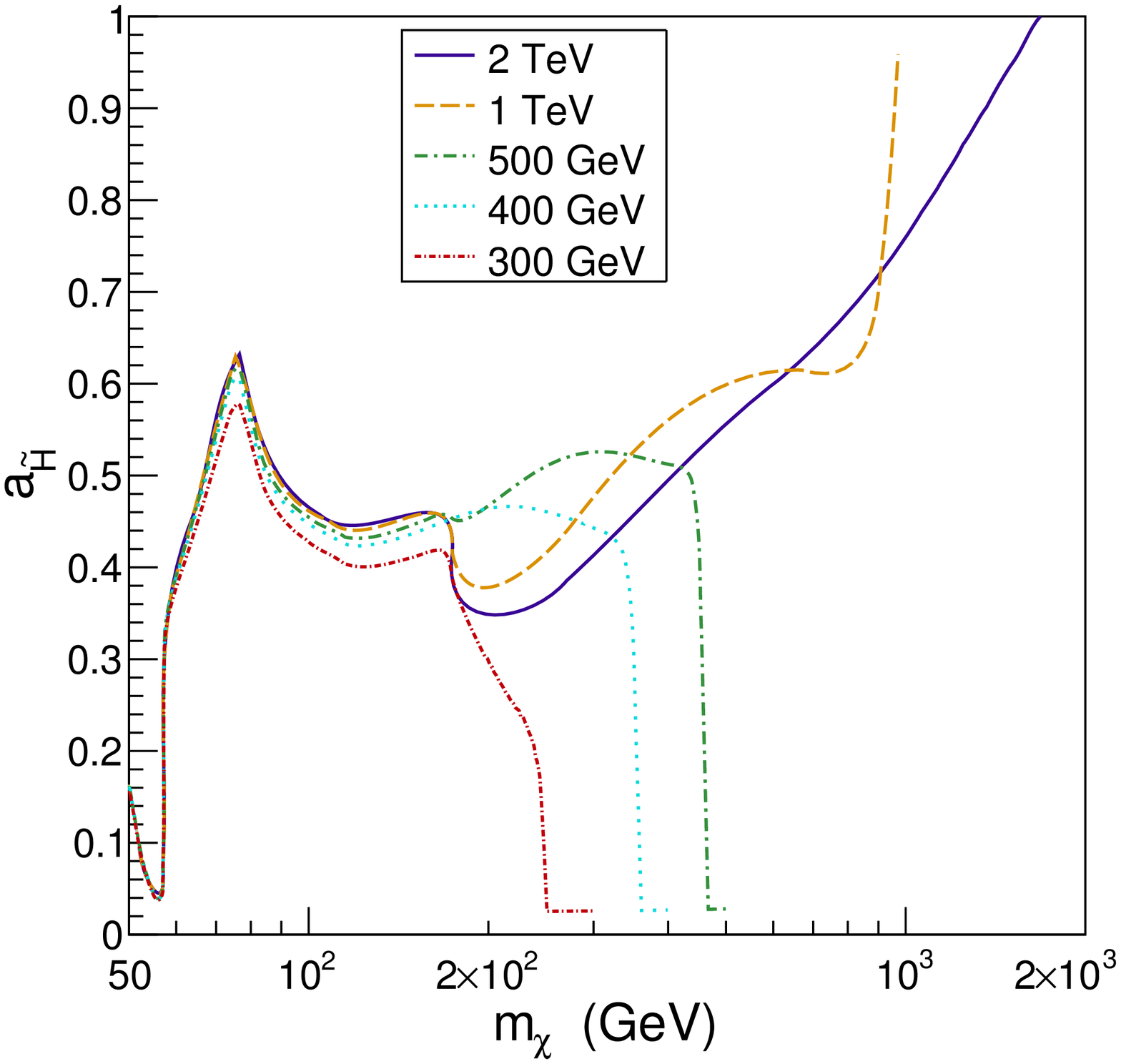}
    \label{fig:ahvarious} }
\caption{{\em Higgsino content of the dark matter particle $\chi$ as a
function of its mass.}  The Higgsino content is fixed by the
requirement $\Omegachi = 0.23$.  Panel \subref{fig:ah2TeV} shows
$\ah$, $|\ahd|$, and $|\ahu|$ for $\mtilde = 2~\tev$ and both choices
of $\signmu$.  The solid line corresponds to $\mu > 0$ and the dotted
line to $\mu < 0$ for $\ah$, and the dashed lines correspond to $\mu >
0$ and dot-dashed lines to $\mu < 0$ for $\ahud$.  Panel
\subref{fig:ahvarious} shows $\ah$ for various values of $\mtilde$ for
$\mu > 0$.  For both panels, we fix $m_A = 4~\tev$ and $\tan\beta =
10$.}
\label{fig:ah}
\end{figure}

In \figref{ah}, we show what Higgsino content is required to satisfy
$\Omegachi = 0.23$ as a function of $\mneut$ for fixed $\signmu$,
$m_A$, and $\tan \beta$.  First consider \figref{ah2TeV}.  We see that
$|\ahu| \sim |\ahd|$ when they are significant, as discussed above.
We also see the general trend that $\ah$ increases with increasing
$\mneut$, because larger $\mneut$ suppresses annihilation and thus
$\ah$ must increase to compensate.  This rise is interrupted three
times: the dip centered at $\mneut \sim 55~\gev$ is caused by the $h$
resonance, where the annihilation cross section is enhanced
kinematically, and so $\ah$ must decrease to compensate dynamically.
In addition, there are two notable drops in $\ah$ at $\mneut \approx
m_W$ and $\mneut \approx m_t$, where annihilation is enhanced by the
opening of a new annihilation channel.  Finally, $\ah$ rises to a $1$
at $\mchi \sim 1.7~\tev$.  This corresponds to a maximum value of
$\mchi$ consistent with the relic density due to decreasing
annihilation efficiency as $\mchi$ increases, barring other effects
(such as co-annihilation) raising the annihilation rate once more.
There is also splitting between the cases of $\mu > 0$ and $\mu < 0$,
a result of the aforementioned dependence of $\ahud$ on the sign of
$\mu$ and the interplay of these parameters in annihilation.

\Figref{ahvarious} shows the dependence of $\ah$ on $\mtilde$. For
$\mneut \le m_t$ the dependence on $\mtilde$ is weak, primarily
because the process $\Neut \Neut \to W^+ W^-$ is independent of
$\mtilde$ and the sfermion-mediated processes are suppressed relative
to this by Yukawa couplings. However, when $\Neut \Neut \to t \bar{t}$
is kinematically allowed, the $\tilde{t}$-mediated process may be
similar in magnitude to the Higgs-mediated process.  In this case, for
$\mtilde = 2~\tev$ the Higgs-mediated annihilation is significantly
larger than the squark contribution and causes a drop in $\ah$ across
the $t \bar{t}$ threshold; for lighter $\mtilde$ the
$\tilde{t}$-mediated contribution cancels the Higgs-mediated
contribution to some degree, and thus less of a drop in $\ah$ is
required.  This effect is especially visible in the case of $\mtilde =
400~\gev$, where there is very little effect from passing $\mneut =
m_t$ corresponding to nearly exact cancellation between the two
processes, and $\mtilde = 300~\gev$, where $\ah$ again drops
significantly because the $\tilde{t}$-mediated diagram is the dominant
contribution.  The curves for lower values of $\mtilde$ also show the
effects of co-annihilation, with the value of $\ah$ dropping rapidly
to a small value then remaining small up to the kinematic
limit.\footnote{These effects should also be present for the 1~TeV and
  2~TeV curves, but were unreachable due to numerical instabilities in
  finding a valid window for the relic density at high $\mchi$.}  The
actual value may vary between zero and a finite but small maximum
value based on the value of $\mu$; we display a value near the maximum
to illustrate later effects upon the scattering cross section and to
minimize the value of $\mu$ in the co-annihilation region at only
$\sim 2~\tev$.

It is instructive to explicitly relate the behavior of $\ah$ in
\figref{ah} back to the fundamental parameters in the theory.
Throughout most of the range of $\mchi$ shown, $0.1 \lesssim \ah
\lesssim 0.9$, which is associated with the focus point region where
$|M_1| \sim |\mu|$ to allow for significant mixing.  In particular,
for that range of $\ah$, $0.67 \lesssim |M_1/\mu| \lesssim 1.05$.  The
significant drops due to the Higgs resonance and co-annihilation
correspond to a nearly pure $\Bino$ LSP with $|M_1| \gg |\mu|$.  We
have set $m_A = 4~\tev$, so there is no $A$-funnel region present in
\figref{ah}, but for a lower value of $m_A$ there would be a small
region of significantly suppressed $\ah$ at $\mchi \sim m_A/2$.  The
presence of an A-funnel effectively obfuscates other interesting
features in the behavior of $\ah$ and ultimately $\sigmaSI$, and this
is part of our motivation for examining a large base value of $m_A$
throughout.  The behavior for lower values of $m_A$ is discussed in
\secref{ma_tanb}.

\section{Spin-Independent Direct Detection}
\label{sec:dd}

\subsection{General Formalism}
\label{sec:sidd}

The relevant interactions for spin-independent scattering are
four-fermion neutralino-nucleon effective operators.  Following
closely the notation of Ref.~\cite{Belanger:2008sj}, these are
\begin{equation}
\label{eq:operator}
\mathcal{L}^{SI} = \lambda_N \bar{\psi}_\chi \psi_\chi \bar{\psi}_N
\psi_N\ ,
\end{equation}
where $N = p, n$, and $\lambda_N$ is determined by the underlying SUSY
interactions that produce the operator.  For a nucleus with charge $Z$
and $A$ total nucleons, this leads to a total cross section of
\begin{equation}
\sigmaSI = \frac{4 \mu_\chi^2}{\pi} \left[ \lambda_p Z + \lambda_n
(A-Z) \right]^2 \ ,
\end{equation}
where $\mu_\chi = \mneut M_A / (\mneut + M_A)$ is the reduced mass of
the $\Neut$-nucleon system.  For neutralinos, typically $\lambda_p
\sim \lambda_n$, and the cross-sections add coherently, so the cross
section can usually be reduced to
\begin{equation}
\sigmaSI \approx \frac{4 \mu_\chi^2}{\pi} \lambda_p^2 A^2 \ .
\end{equation}

\begin{figure}[tbp]
\begin{center}
 \unitlength = 0.75mm
 \subfigure[Higgs-mediated Scattering]{
   \includegraphics[width=.3\textwidth]{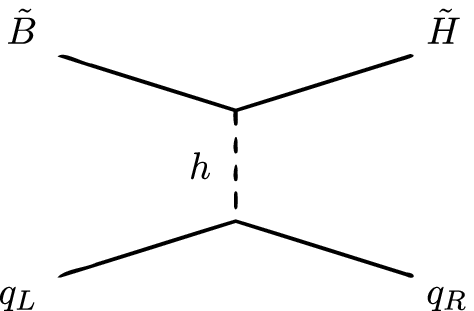}
   \label{fig:scatteringHiggs} }
 \subfigure[Squark-mediated Scattering]{
   \includegraphics[width=.3\textwidth]{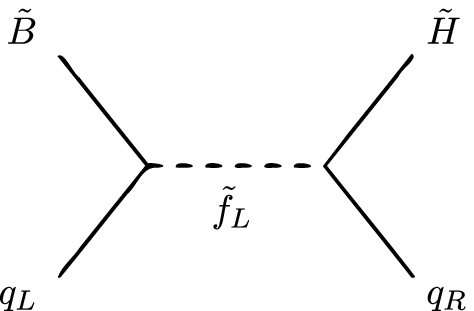}
   \label{fig:scatteringQ} }
 \subfigure[Squark-mediated Scattering with LR Mixing]{
   \includegraphics[width=.3\textwidth]{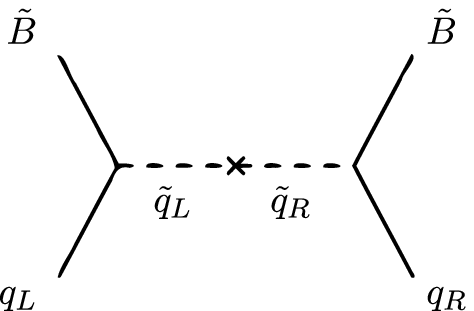}
   \label{fig:scatteringQLR} }
\end{center}
\caption{{\em Major processes contributing to $\sigmaSI$.}  The
diagrams are labeled with the appropriate gauge eigenstates, and
additional diagrams exist with the exchange $\Bino \leftrightarrow
\Higgsino$ and either $f_L \leftrightarrow f_R$ or $\tilde{f}_L
\leftrightarrow \tilde{f}_R$.  This exchange is independent in
\subref{fig:scatteringHiggs} and correlated in
\subref{fig:scatteringQ} and \subref{fig:scatteringQLR}.}
\label{fig:scattering}
\end{figure}

The dominant diagrams contributing to $\sigmaSI$ for mixed
$\Bino-\Higgsino$ are shown in \figref{scattering}, all of which have
different coupling strengths for different quark flavors.  Because of
this, the cross-section has a strong dependence on the individual
contributions of each quark flavor to the current $\bar{\psi}_q
\psi_q$ in the nucleon.  This quantity is parameterized as
\begin{equation}
\left\langle N \left| m_q \bar{\psi}_q \psi_q \right| N \right\rangle
= f_q^N M_N \ ,
\end{equation}
where $M_N$ is the nucleon mass and $f_q^N$ is interpreted as the
contribution of the quark $q$ to the nucleon mass.

Using this parameterization, the coupling of the neutralino to the
nucleon becomes
\begin{equation}
\frac{\lambda_N}{m_N} = \sum\limits_{q = 1}^6 f_q^N
\frac{\lambda_q}{m_q} \ ,
\end{equation}
where $\lambda_q$ is the effective coupling of each individual quark
to the neutralino through an operator analogous to \eqref{eq:operator}
above with $N \to q$.  For the light quarks $u$, $d$, and $s$, the
values of $f_q^N$ can be taken from meson scattering experiments or
lattice results.  For the heavy quarks $Q = c, b, t$, the entire
contribution derives from loop contributions through gluon exchange,
and has the value~\cite{Shifman:1978zn}
\begin{equation}
f_Q^N = \frac{2}{27} \left( 1 - \sum\limits_{q = u, d, s} f_q^N
\right) \ .
\end{equation}
The parameters $f_{u,d}^N$ are reasonably well-known and small, but
$f_s^N$ has been the subject of debate due to varying experimental and
lattice results.  In our analysis we take the values of $f_{u,d}^N$
from Ref.~\cite{Belanger:2008sj} and use the lattice result given in
Ref.~\cite{Giedt:2009mr} for $f_s^N$, which gives
\begin{eqnarray}
&& f_d^p = 0.033\, , \ f_u^p = 0.023\, , \ f_s^p = 0.05 \\
&& f_d^n = 0.042\, , \ f_u^n = 0.018\, , \ f_s^n = 0.05 \ .
\end{eqnarray}
These values imply $f_Q^p = f_Q^n = 0.066$, so the heavy quark
contribution is quite important to the overall value of $\sigmaSI$. We
will discuss the impact of variations from these values in
\secref{fs}.

\subsection{Neutralino-Quark Couplings}
\label{sec:lambdaq}

The coefficients for spin-independent scattering of neutralinos off
individual quarks, $\lambda_q$, have the form~\cite{Ellis:2000jd}
\begin{eqnarray}
\nonumber \lambda_u &=& - \frac{g^2}{2 \left(m_{\tilde{u}_i}^2 -
  \mneut^2 \right)} \left[ \left( \frac{1}{2} \aw + \frac{1}{6}
\tan\theta_W \ab\right) U^{(\tilde{u})*}_{i1} + \frac{m_u \ahu}{2 m_W
  \sin\beta} U^{(\tilde{u})*}_{i2} \right] \\
\nonumber & & \hspace*{5cm} \times \left( \frac{m_u \ahu}{2 m_W
\sin\beta} U^{(\tilde{u})}_{i1} - \frac{2}{3} \tan\theta_W \ab
U^{(\tilde{u})}_{i2} \right) \\
\nonumber & & - \frac{g^2 m_u}{4 m_W \sin\beta} \left(\aw -
\tan\theta_W \ab \right) \left[ \left( \ahd \cos\alpha \sin\alpha +
\ahu \cos^2 \alpha\ \right) \frac{1}{m_h^2} \right. \\
& & \hspace*{5.5cm} \left. + \left( - \ahd \cos\alpha \sin\alpha + \ahu
  \sin^2 \alpha\ \right) \frac{1}{m_H^2} \right] \label{eqn:lambdau} \\
\nonumber \lambda_d &=& - \frac{g^2}{2 \left(m_{\tilde{d}_i}^2 -
  \mneut^2 \right)} \left[ \left( - \frac{1}{2} \aw + \frac{1}{6}
  \tan\theta_W \ab \right) U^{(\tilde{d})*}_{i1} + \frac{m_d \ahd}{2
  m_W \cos\beta} U^{(\tilde{d})*}_{i2} \right] \\
\nonumber & & \hspace*{5cm} \times \left( \frac{m_d \ahd}{2 m_W
\cos\beta} U^{(\tilde{d})}_{i1} + \frac{1}{3} \tan\theta_W \ab
U^{(\tilde{d})}_{i2} \right) \\
\nonumber & & + \frac{g^2 m_d}{4 m_W \cos\beta} \left(\aw -
\tan\theta_W \ab\right) \left[ \left( \ahu \cos\alpha \sin\alpha +
\ahd \sin^2 \alpha\ \right) \frac{1}{m_h^2} \right. \\
& & \hspace*{5.5cm} + \left. \left( - \ahu \cos\alpha \sin\alpha + \ahd
  \cos^2 \alpha\ \right) \frac{1}{m_H^2} \right] , \label{eqn:lambdad}
\end{eqnarray}
where $\alpha$ is the CP-even Higgs mixing angle, $m_{\tilde{q}_i}$
and $U^{(\tilde{q})}_{ij}$ are the squark masses and mixing matrix
defined by
\begin{equation}
M_{\tilde{q}}^2 = U^{(\tilde{q})\dagger} \left(
\begin{array}{cc}
m_{\tilde{q}_1}^2 & 0 \\
0 & m_{\tilde{q}_2}^2
\end{array} \right) U^{(\tilde{q})} \ ,
\end{equation}
and there is an implicit sum over $i$.  The $m_{h,H}$-dependent terms
of \eqsref{eqn:lambdau}{eqn:lambdad} arise from the diagram of
\figref{scatteringHiggs}, and the $m_{\tilde{q}}$-dependent terms are
from the diagrams of \figref{scatteringQ} and \figref{scatteringQLR}.
The coupling for heavier quarks can be determined by simple mass
replacement in the appropriate equation.  These couplings can be altered
somewhat by the inclusion of an additional CP-violating phase in the
Higgs sector~\cite{Falk:1999mq}, but we will not address this
contribution in this work.

We have assumed $M_{\tilde{q}}^2 = \mtilde^2 \mathds{1}$, which
implies $U^{(\tilde{q})} = \mathds{1}$.  It is instructive to consider
the limit $m_H \sim m_A \to \infty$.  In this case, $\alpha \sim \beta
- \pi/2 < 0$, and the couplings simplify to
\begin{eqnarray}
\label{eqn:lambdaud}
\lambda_u & = & - \frac{g^2 m_u \left(\aw - \tan\theta_W \ab
  \right)}{4 m_W \sin\beta} \left[ \frac{\ahu}{\mtilde^2 - \mneut^2} +
  \frac{ - \ahd \sin\beta \cos\beta + \ahu \sin^2 \beta}{m_h^2}
  \right] \\
\lambda_d & = & \frac{g^2 m_d \left(\aw - \tan\theta_W \ab
  \right)}{4 m_W \cos\beta} \left[ \frac{\ahd}{\mtilde^2 - \mneut^2} +
  \frac{ - \ahu \sin\beta \cos\beta + \ahd \cos^2 \beta}{m_h^2}
  \right] \ .
\end{eqnarray}
The dependence of $\lambda_q$ on the Higgsino-ness of the neutralino
is manifest, with the coupling vanishing for a pure $\Bino$ or
$\Higgsino$ and the maximum value generically achieved for a
well-mixed case.  Moreover, it is clear that the squark-mediated
diagrams are sub-dominant relative to the Higgs-mediated diagrams,
except when $m_{\tilde{d}}^2 - \mneut^2 \sim m_h^2$, in which case
their contributions become comparable.

Assuming the Higgs-mediated contributions dominate, there is also
clearly a dependence on the relative signs of $\ahu$ and $\ahd$, both
in the Higgs-mediated contribution alone and in the relative sign
between the contributions.  This contribution is either a suppression
or enhancement in both types of quarks at once, with a characteristic
suppression of the couplings when $\ahud$ have the same sign relative
to the values with opposite signs.  Moreover, in the same-sign case
there is the possibility of cancellation between the two contributions
further reducing $\sigmaSI$, while for the opposite-sign case there is
an enhancement.  \Eqref{eqn:ahud} implies that $\ahu$ and $\ahd$ will
have the same sign for $\mu < 0$ and opposite signs for $\mu > 0$.
This means there will be a characteristic enhancement in the $\mu > 0$
value of $\sigmaSI$ over the $\mu < 0$ value for a given set of other
parameters.  This holds true over most of our parameter space.  The
most significant exception is when the LSP is a nearly pure Bino and
\eqref{eqn:ahud} doesn't hold.  In this case, $\ah \ll 1$, which
causes a severe suppression in $\lambda_q$ and thus in $\sigmaSI$ as
well, regardless of the sign of $\mu$.

\section{Characteristic Spin-Independent Cross Sections}
\label{sec:main}

\begin{figure}[tbp]
\includegraphics[width=0.72\textwidth]{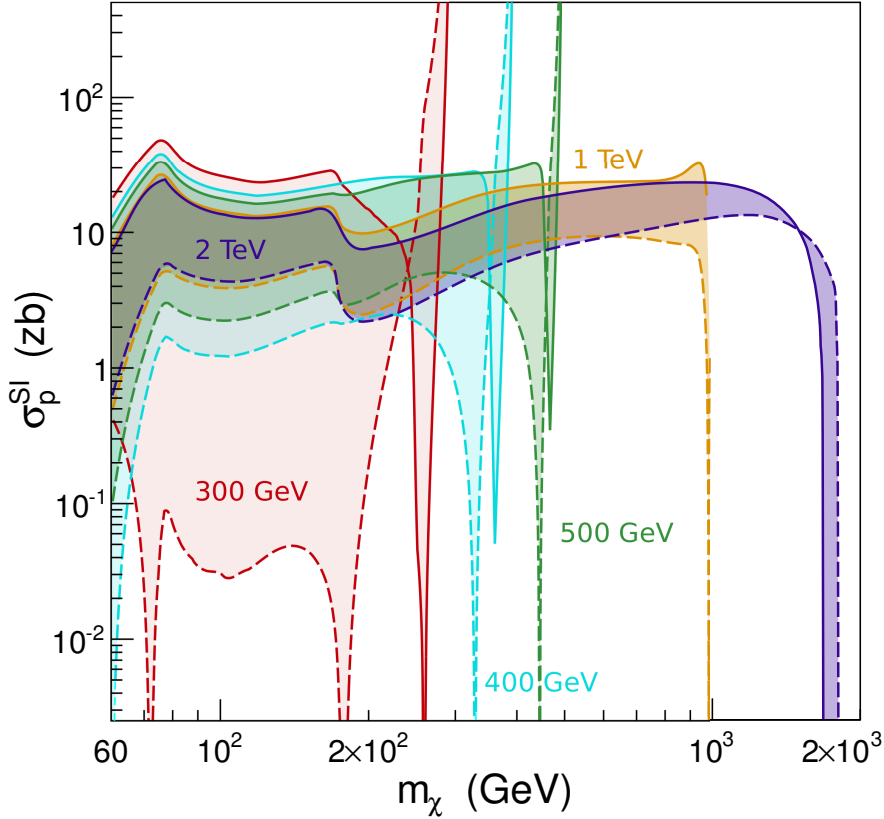}
\caption{{\em Spin-independent neutralino-nucleon cross sections.} The
cross sections $\sigmaSI$ for models with thermal relic density
$\Omegachi = 0.23$ as functions of neutralino mass $\mneut$ for the
various sfermion masses indicated, $\mu > 0$ (solid) and $\mu < 0$
(dashed), $m_A = 4~\tev$ and $\tan\beta = 10$.  The shaded region
between the solid and dashed curves are obtained for complex values of
$\mu$ that interpolate between the two real values (see
\secref{cpinclusion}).}
\label{fig:sigmas}
\end{figure}

We are now ready to present our main results for the spin-independent
neutralino-nucleon cross sections in our model framework.  In
\figref{sigmas}, we present the predicted values for $\sigmaSI$ as a
function of $\mneut$ for models with the correct thermal relic density
$\Omegachi = 0.23$.  Results are given for various $\mtilde$, both
signs of $\mu$, $m_A = 4~\tev$, and $\tan\beta = 10$.  The dependence
on $m_A$ and $\tan\beta$ is examined in \secref{ma_tanb}.

\subsection{Characteristic Features}

\Figref{sigmas} has a number of striking characteristic features.
First, aside from the results for $\mtilde=300~\gev$ and $\mu <0$,
which we discuss below, the predicted cross sections $\sigmaSI$ are
almost independent of dark matter mass.  For $\mtilde = 2~\tev$, this
may be understood as follows: in this case, sfermions are decoupled,
and scattering occurs through the light Higgs boson.  These cross
sections scale as $\sim \ah^2/m_h^4$, but the Higgsino-ness is
typically constrained by the thermal relic density to the range $0.3
\alt \ah \alt 0.8$, as shown in \figref{ah}, and the Higgs boson mass
is constrained to the narrow range $114.4~\gev < m_h \alt 130~\gev$,
given the current experimental bound.  As a result, for $\mtilde =
2~\tev$, $m_A = 4~\tev$ and $\tan \beta = 10$, $\sigmaSI$ is in the
narrow range $2~\zb < \sigmaSI < 20~\zb$ for all $\mneut \agt
70~\gev$.

Perhaps more surprising is that this behavior persists to much lower
sfermion masses.  As evident in \figref{sigmas}, even for $\mtilde =
400~\gev$, the characteristic range of $\sigmaSI$ is extended only to
$1~\zb < \sigmaSI < 40~\zb$ for $\mneut \agt 70~\gev$.  Such low
values of $\mtilde$ are not far from current bounds on squark masses
from direct searches at the Tevatron~\cite{Feng:2009te}.  In this
sense, current bounds from colliders already imply that sfermions are
essentially decoupled from neutralino phenomenology, except in
fine-tuned cases discussed below, greatly simplifying SUSY parameter
space and focusing predictions.

In more detail, we also see that the $\sigmaSI$ curves of
\figref{sigmas} show a generic similarity to the $\ah$ curves of
\figref{ah} as suggested by the form of the coefficients $\lambda_q$.
There is a peak and subsequent drop-off in the distribution at $\mneut
\sim m_W$ and again at $\mneut \sim m_t$ corresponding to the
fluctuations in the $\ah$, and most of the curves demonstrate some
rise after $\mneut \sim m_t$.  For $\mtilde = 2~\tev$, $\sigmaSI$
drops of at $\mchi \lesssim 1.7~\tev$ as the neutralino becomes nearly
pure $\Higgsino$.  This because, in the absence of left-right mixing,
the dominant scattering processes require $\chi$ to have both
significant $\Bino$ and $\Higgsino$ components as shown in
\figref{scattering}.

Finally, for the reasons discussed in \secref{lambdaq}, the value of
$\sigmaSI$ for $\mu < 0$ is almost always suppressed relative to that
for $\mu > 0$, usually by at least a factor of two.  This effect is
enhanced for lower $\mtilde$.  This follows directly from the form of
$\lambda_q$ above, with decreasing scalar masses bringing greater
enhancement for $\mu>0$ and greater suppression for $\mu < 0$.
Although this relation is not strictly true, due to the more
pronounced dip in $\sigmaSI|_{\mu<0}$ for larger $\mtilde$ at $\mneut
\sim m_t$, the suppression in $\sigmaSI$ for $\mtilde = 2~\tev$ is
relatively minor.

\subsection{Fine-tuned Cancellations}

The cross section $\sigmaSI$ may move outside its characteristic range
in certain fine-tuned regions.  Given our simple, phenomenological
model framework, we may identify the fine-tuned possibilities
relatively easily.

First, for $50~\gev \alt \mneut \alt 65~\gev$, $\sigmaSI$ may be
suppressed by orders of magnitude, consistent with the suppression of
$\ah$ shown in \figref{ah}.  This behavior is visible at the left-hand
edge of \figref{sigmas}.  For these masses, $m_h \approx 2 \mneut$ to
within a few GeV, $\Neut \Neut$ annihilation is enhanced by the Higgs
boson resonance, and so the correct thermal relic density is obtained
even for Bino-like neutralinos.  In this fine-tuned case, $\sigmaSI$
may be so small that neutralinos will escape all direct detection
searches for the foreseeable future.  Note, however, that this
possibility is disfavored not only by fine-tuning, but will soon be
probed by chargino and gluino searches, since these are sensitive to
LSP masses around 50 GeV under the assumption of gaugino mass
unification.

The regions in \figref{sigmas} must, of course, be cut off at $\mneut
= \mtilde$; for $\mneut > \mtilde$, the neutralino is not the LSP.  In
fact, the behavior of $\sigmaSI$ changes significantly for values
$\sim 20-50~\gev$ below $\mtilde$.  The reason for this is that the
annihilation is enhanced for $\mneut \sim \mtilde$ by co-annihilation
effects, quickly suppressing the relic density once co-annihilation
becomes important.  Although $\ah$ may compensate for this effect, the
$\ah$ curves in \figref{ahvarious} demonstrate rapid change as
$\mneut$ approaches the end of the curve, indicating that $\Neut$ will
quickly become a pure $\Bino$ or $\Higgsino$ to suppress annihilation
somewhat before $\mneut = \mtilde$.  After this point the correct
relic density is unattainable.  However, in contrast to typical
assumptions regarding co-annihilation, the value of $\sigmaSI$ may
actually increases greatly approaching the kinematic limit.  This is
due to the unified sfermion mass, which results in co-annihilation
with squarks as well as sleptons, and thus there is a near-resonance
enhancement of $\sigmaSI$ if $\ah$ remains small but constant.  This
effect is largely illustrative, as most realistic scenarios predict
slepton masses at least somewhat smaller than squark masses, and the
exact values of $\sigmaSI$ are undetermined as the exact value of
$\ah$ is not set in the co-annihilation region.

For $\mtilde = 1~\tev$, $500~\gev$, and $400~\gev$, there is a small
region with a significantly suppressed value of $\sigmaSI$ in the $\mu
< 0$ curves for $\mneut$ just below the co-annihilation cutoff.  To
understand this feature, consider the form of $\lambda_d$ from
\eqref{eqn:lambdaud}.  For large $\tan\beta$, $\sin \alpha$ is small
and negative, so $\lambda_d = 0$ for
\begin{equation}
m_{\tilde{d}}^2 - \mneut^2 \sim -
\frac{\ahd}{\ahu \cos \alpha \sin \alpha} m_h^2
\sim \frac{2\mu}{\mneut \sin 2\alpha} m_h^2 \sim 10 m_h^2
\end{equation}
for characteristic values of $\alpha$ and $\mu/\mneut$.  For even
moderately larger values of $\mtilde$, the $\tilde{q}$-mediated
contribution is sub-dominant until $\mneut$ becomes sufficiently
large.  For $\mneut^2 \sim m_{\tilde{d}}^2 - 10 m_h^2$, $\lambda_d$
vanishes at a specific value then flips sign for larger values of
$\mneut$.  For $\lambda_u = 0$, the corresponding condition is
\begin{equation}
m_{\tilde{u}}^2 - \mneut^2 \sim - \frac{1}{\cos^2 \alpha -
\frac{\mu}{\mneut} \cos \alpha \sin \alpha} m_h^2 \ ,
\end{equation}
so $\lambda_u$ cannot vanish for $\mneut < \mtilde$.  The highly
suppressed region is the result of cancellation between contributions
from up-type and down-type quarks in this region, the result of the
squark-mediated contributions to $\sigmaSI$ in \figref{scattering}
becoming similar in size to the Higgs-mediated contribution for
$\mneut \sim \mtilde$.

In fact, this effect explains the suppression of $\sigmaSI$ for
$\mtilde = 300~\gev$ across a large range of $\mneut$.  In this case
$\mtilde^2 \sim 10 m_h^2$ regardless of the value of $\mneut$, so the
Higgs- and squark-mediated contributions are always comparable.  This
leads to a general suppression in $\sigmaSI$ until $\mneut \sim
\mtilde$, where $\sigmaSI$ actually increases to the characteristic
value because the squark-mediated contribution dominates.

\subsection{Prospects for Discovery}

The primary implication of \figref{sigmas} is that simple SUSY models
typically predict $1~\zb \alt \sigmaSI \alt 40~\zb$.  This prediction
becomes even more narrow for large sfermion masses, which are required
for ${\cal O}(1)$ phases to be consistent with the EDM constraints
discussed in \secref{cp}. The reason lies in the decoupling of
squark-mediated processes in \figref{scatteringQ} and
\figref{scatteringQLR}, leaving only the Higgs mediated process in
\figref{scatteringHiggs}.

\begin{figure}
\includegraphics[width=0.72\textwidth]{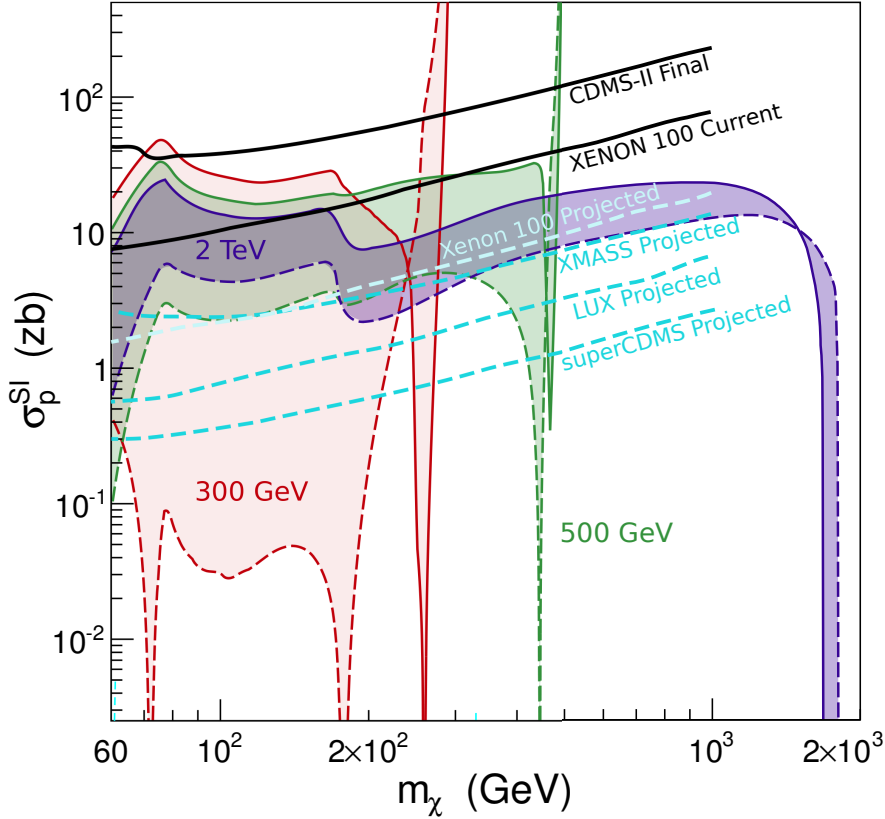}
\caption{{\em Theoretical predictions for $\sigmaSI$ and experimental
    bounds.}  The shaded regions are as in \figref{sigmas}, along with
  various current and project bounds.  The current upper limit from
  CDMS~\cite{Ahmed:2009zw} skirts our projected cross sections, and
  recent XENON100~\cite{Aprile:2011hi} results have begun to constrain
  our characteristic region.  Most or all of the characteristic
  regions will be probed by the projected near future sensitivities of
  XENON100~\cite{Aprile:2009yh}, XMASS~\cite{Sekiya:2010bf},
  LUX~\cite{Fiorucci:2009ak,LUXwebsite}, and superCDMS with 100 kg
  fiducial mass at SNOLAB~\cite{Bruch:2010eq}.}
\label{fig:sigmas_limits}
\end{figure}

Such cross sections are just below current bounds for most masses, and
recent XENON100~\cite{Aprile:2011hi} results have begun probing our
characteristic region.  These predictions along with current
experimental bounds and expected near future sensitivities, are shown
in \figref{sigmas_limits}.  We see that the expected range is just
below current limits, but the characteristic range will be probed by
experiments that reach the zeptobarn scale in the near future.

\section{Dependence on Model Parameters and Assumptions}
\label{sec:caveats}

In this section we examine the effects of relaxing several of our
particle physics model assumptions, varying the strange quark content
of the nucleon, and modifying astrophysical inputs that enter our
analysis.

\subsection{Variations in $m_A$ and $\tan\beta$ }
\label{sec:ma_tanb}

\begin{figure}[tbp]
  \subfigure[Variation of $m_A$]{
    \includegraphics[width=.48\textwidth]{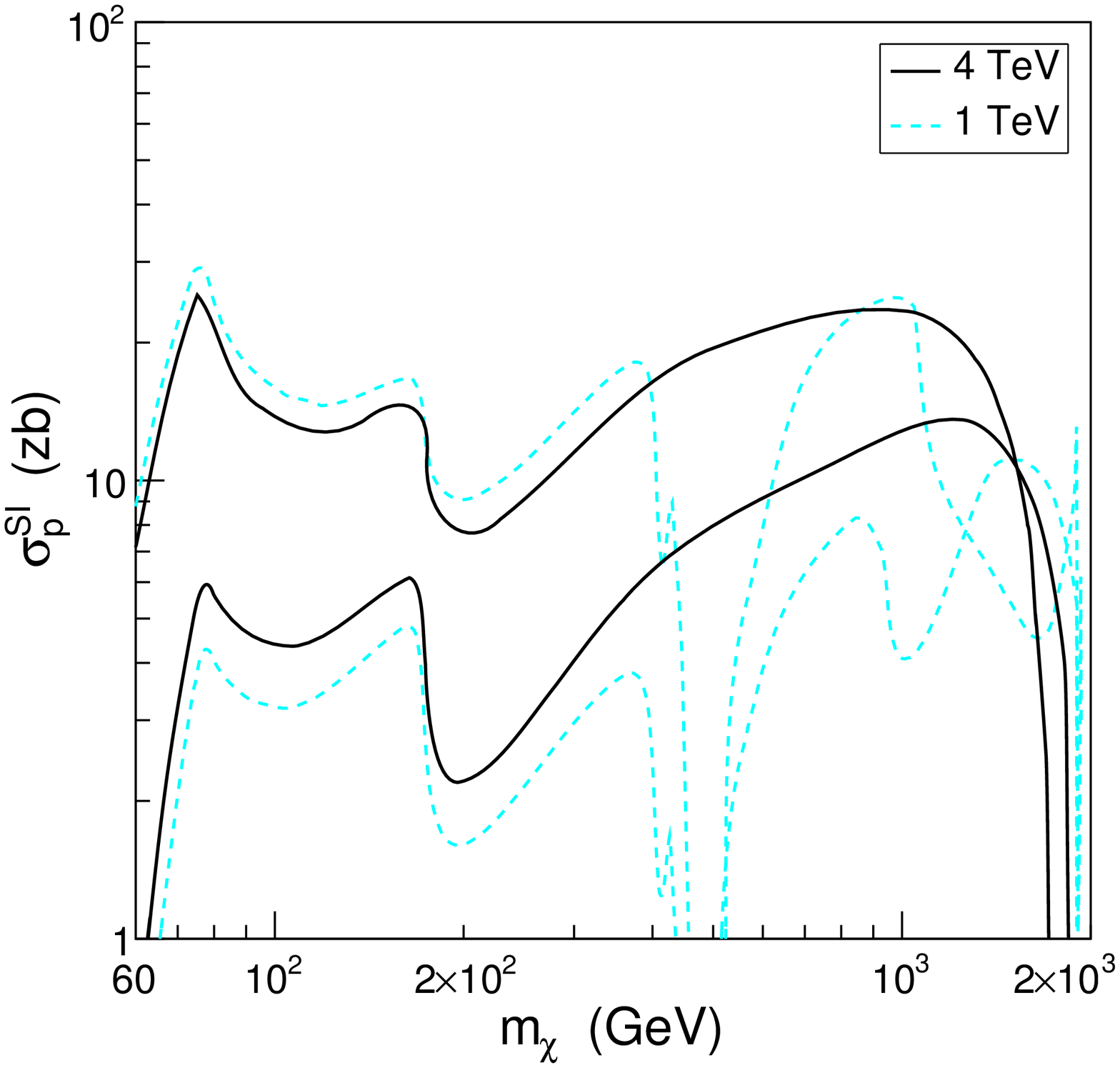}
    \label{fig:low_mA} }
  \subfigure[Variation of $\tan\beta$]{
    \includegraphics[width=.48\textwidth]{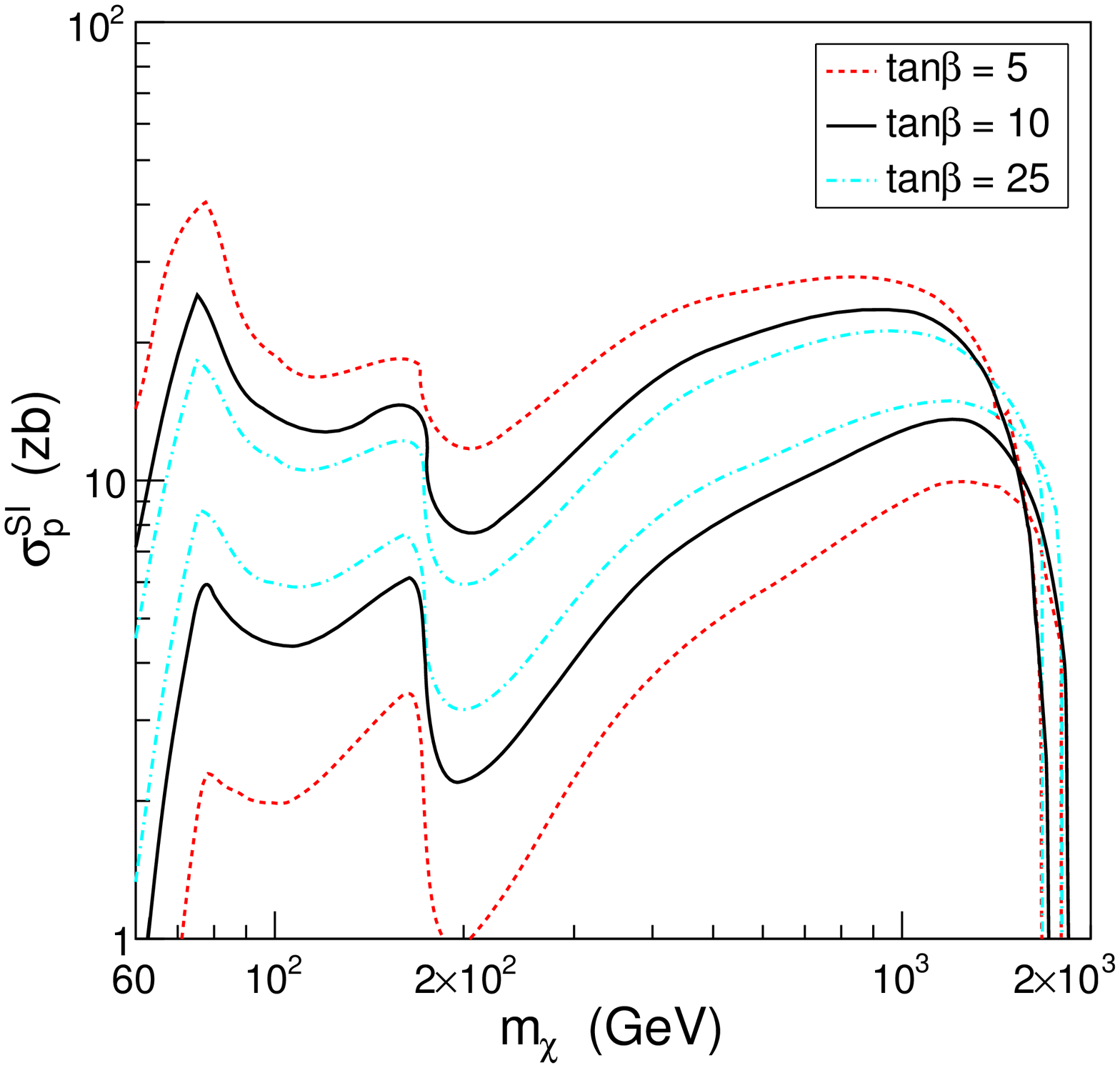}
    \label{fig:tanbeta} }
\caption{{\em Dependence of $\sigmaSI$ on Higgs potential parameters.}
The cross section $\sigmaSI$ is given for various $m_A$ and
$\tan\beta=10$ in \subref{fig:low_mA} and various $\tan\beta$ and $m_A
= 4~\tev$ in \subref{fig:tanbeta}.  In all cases, $\mtilde = 2~\tev$,
and the upper contour is for $\mu >0$ and the lower contour is for
$\mu<0$.}
\label{fig:higgsvariation}
\end{figure}

As discussed in \secref{model}, our Higgs potential is parameterized
by $\mu$, $m_A$, and $\tan\beta$.  After imposing the constraint
$\Omegachi = 0.23$, the magnitude of $\mu$ is determined, and we have
discussed the dependence of our results on the choice of $\signmu$.
\Figref{higgsvariation} shows the effects of varying the remaining
input parameters of the Higgs potential: $m_A$ and $\tan \beta$.

As noted in \secref{extrapolation}, overly efficient annihilation and
stability of the Higgs potential at high scales generally motivate at
least partially decoupled values of $m_A \agt 300 - 500~\gev$.  Up to
this point in our analysis, we have considered the complete decoupling
limit with $m_A = 4~\tev$.  Here we consider the effects of a lower
value of $m_A$ by showing results for $m_A = 1~\tev$ in
\figref{low_mA}.  The primary effect is a strong suppression of
$\sigmaSI$ for $\mneut \sim m_A /2$.  For such values of $\mneut$,
$\Neut \Neut$ annihilation is enhanced by the $S$-wave $A$ resonance.
This effect is similar to the light Higgs resonance discussed
previously, as it forces $\Neut$ to be nearly pure $\Bino$ to
compensate for the strong annihilation enhancement and thus suppressed
$\sigmaSI$.  This behavior is present in the $A$-funnel region at
large $\tan\beta$ in mSUGRA, but we see here that it is generically
possible for all values of $\tan\beta$ and the position in parameter
space is generically unconnected to the sfermion masses. There is also
a slightly less pronounced resonance involving the heavier CP-even
Higgs which produces the less distinct second downward spike for the
$\mu < 0$ curve.  The value of $\sigmaSI$ drops somewhat for $\mchi >
1~\tev$ due to the availablitiy of heavy Higgs final states in
annihilation.  The availability of $AA$ final states also increases
the maximum value of $m_\chi$ by increasing annihilation efficiency.
Finally, there is a slight broadening of the values of $\sigmaSI$ due
to the increased contribution from scattering through the heavy
CP-even Higgs; this effect will become more pronounced for even lower
$m_A$, but the aforementioned lower limit on $m_A$ prevents this
effect from causing serious suppression.

For variation in $\tan\beta$, the characteristic range of values for
$\sigmaSI$ narrows for larger values of $\tan\beta$.  This can be
understood as follows: for the neutralino mass matrix, in the limit
$\tan\beta \to \infty$, the sign of $\mu$ may be switched by the field
redefinition $\Higgsino_d \to - \Higgsino_d$ without affecting other
matrix entries.  We therefore expect the $\mu > 0$ and $\mu < 0$
predictions to become similar for large $\tan\beta$.  Conversely, as
shown in \figref{tanbeta}, the range of $\sigmaSI$ becomes wider for
lower values of $\tan\beta$; for $\tan\beta = 5$, roughly the minimum
value allowed by null results from Higgs boson searches, the lowest
value of $\sigmaSI$ is reduced by a factor $\sim 2$ from the
$\tan\beta = 10$ results.

For very large $\tan \beta$ and low value of $m_A$, there is a
possibility of a general suppression due to cancellation between light
and heavy Higgs contributions to $\sigmaSI$.  For $\tan \beta
\rightarrow \infty$, $\sin \alpha \rightarrow 0$ and $\cos \alpha
\rightarrow 1$.  Then, if the squark mediated contributions are
neglected, the couplings in \eqsref{eqn:lambdau}{eqn:lambdad} become
\begin{eqnarray}
\lambda_u &\simeq& - \frac{g^2 m_u}{4 m_W \sin\beta} \left(\aw -
\tan\theta_W \ab \right) \frac{\ahu}{m_h^2} \\
\lambda_d &\simeq& \frac{g^2 m_d}{4 m_W \cos\beta} \left(\aw -
\tan\theta_W \ab\right) \frac{\ahd}{m_H^2}\ .
\end{eqnarray}
Then, under the (very) rough assumption of equal portion of up- and
down-type quarks in the nucleon, there is a cancellation between these
contributions when $\lambda_u/m_u + \lambda_d/m_d \simeq 0$, which
gives
\begin{equation}
\frac{m_H}{m_h} \simeq \sqrt{\frac{\ahd}{\ahu} \tan\beta} \simeq
\sqrt{- \frac{\mu}{\mchi} \tan\beta}
\end{equation}
This allows for a possible cancellation for $\mu < 0$ with little
dependence on $\mchi$.  For $\tan\beta = 10$ the required value is
$m_H \simeq m_A \sim 300-400~\gev$, which strains the relic density
calculation; however, for $\tan\beta = 50$, the cancellation occurs at
larger values of $m_A \sim 700-800~\gev$.  However, even for large
$\tan\beta$ significant suppression only occurs at very specific
values of $m_A$, as for lower values of $m_A$ the contribution to
scattering from the heavy Higgs dominates the cross section.

\subsection{Un-unified Sfermion Masses}
\label{sec:splitmasses}

Although convenient for studying the general behavior of $\sigmaSI$,
our assumption of a unified weak-scale sfermion mass is not motivated
by a connection to a high-energy theory.  Indeed, most high energy
theories predict $m_{\slepton} < m_{\tilde{q}}$, either resulting from
a unified scalar mass at the GUT scale, as in mSUGRA, or as a remnant
of sfermion masses generated by gauge dynamics, as in gauge-mediated
supersymmetry breaking models or anomaly-mediated supersymmetry
breaking models with a unified scalar mass.

\begin{figure}[tbp]
  \subfigure[Variation in $m_{\slepton}$]{
    \includegraphics[width=.48\textwidth]{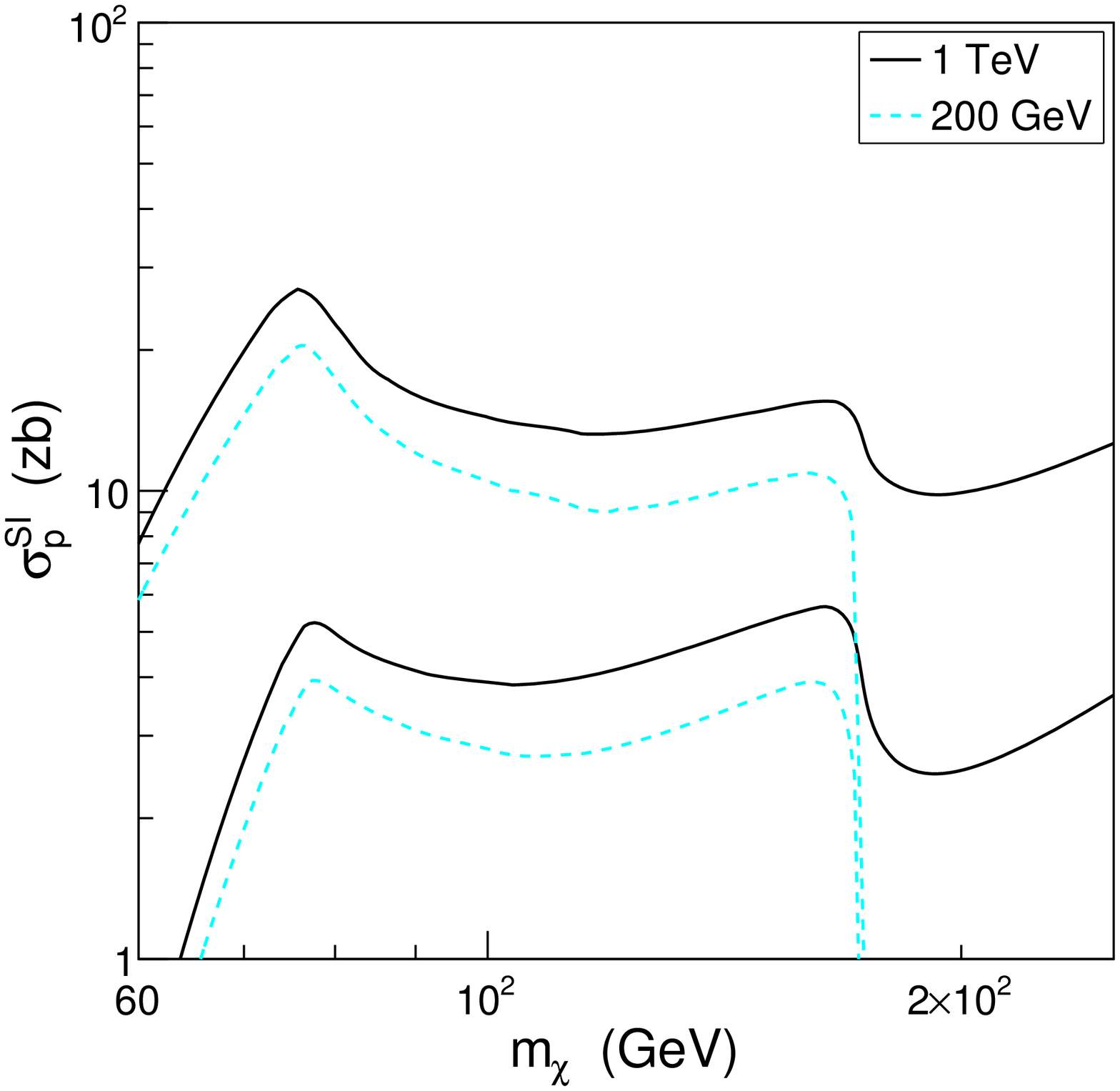}
    \label{fig:lowslep} }
  \subfigure[Variation in $m_{\tilde{t}, \tilde{b}}$]{
    \includegraphics[width=.48\textwidth]{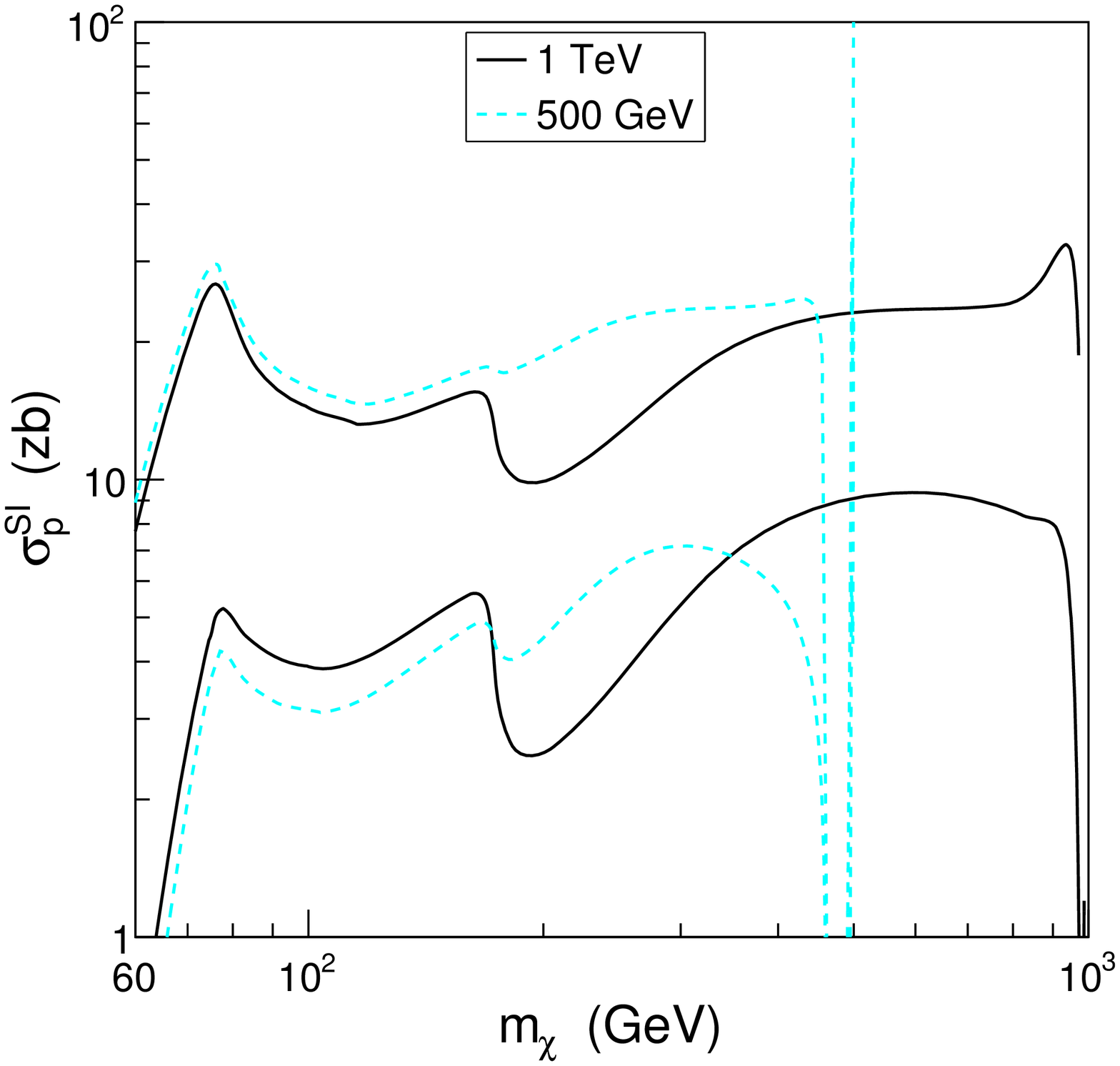}
    \label{fig:lowthird} }
\caption{{\em Dependence of $\sigmaSI$ on the unified sfermion mass
assumption.}  Cross sections $\sigmaSI$ are given for $\mu > 0$
(upper) and $\mu < 0$ (lower), $m_A = 4~\tev$, and $\tan\beta = 10$.
Sfermions masses that are not specified are set to $\mtilde =
1~\tev$.}
\label{fig:lowmasses}
\end{figure}

To explore the consequences of lighter sleptons, in \figref{lowslep},
we compare our standard scenario with $\mtilde = 1~\tev$ to a scenario
with a unified weak-scale squark mass $m_{\tilde{q}} = 1~\tev$ and a
separate unified weak-scale slepton mass $m_{\slepton} = 200~\gev$.
The lower value for $m_{\slepton}$ reduces the value of $\sigmaSI$
across the entire range of masses by $\sim 20-30\%$.  This is due to
the effect of $t$-channel $\slepton$-mediated diagrams in $\Neut$
annihilation reducing $\ah$ and providing no additional to $\sigmaSI$.
As expected, the spike of high $\sigmaSI$ associated with squark
co-annihilation in \figref{sigmas} figures is absent.  While it is
possible that even smaller values of $m_{\slepton}$ could produce much
stronger suppression of $\sigmaSI$ over a large range of $\mneut$
(that is, not just in co-annihilation regions), the value of $
m_{\slepton} = 200~\gev$ is already in significant tension with
electron EDM constraints.  Thus we conclude that the main effect of
lower $m_{\slepton}$ is to reduce the range of $\mneut$ for which the
$\Neut$ can be the LSP, except perhaps for very low values
$m_{\slepton}$ which imply significant fine-tuning in CP phases.

Another well-motivated non-unified sfermion mass possibility is to
have heavy first and second generation sfermions and light third
generation sfermions.  Precision measurements strongly constrain the
mixing in the first two generations, but the mixing between the first
two generations and the third generation is less well-constrained.  An
interesting possibility, then, is that the squarks of the first two
generations are heavy to limit contributions to low-energy observables
through decoupling, while the third generation squark masses are light
to avoid fine-tuning in the electroweak
scale~\cite{Dimopoulos:1995mi,Dvali:1996rj,Cohen:1996vb,Feng:1998iq,%
Bagger:1999ty}.  The impact of such a sfermion mass pattern is shown
in \figref{lowthird}, which contrasts results for $\mtilde = 1~\tev$
and the case where $m_{\tilde{t},\tilde{b}} = 500~\gev$, but all other
sfermions remain at 1 TeV.  Beyond lowering the upper limit on
$\mneut$ such that $\Neut$ remains the LSP, a lower value for
$m_{\tilde{t}}$ reduces the suppression of $\sigmaSI$ associated with
$t \bar{t}$ annihilation, as discussed in \secref{dd}.  There is also
a slight widening of allowed values of $\sigmaSI$, as $\lambda_{b,t}$
do contribute to $\sigmaSI$, but this widening is not as significant
as that caused by lowering $\mtilde$.  Overall, the effect is larger
than the effect of varying slepton masses, but again, it is not very
significant.

\subsection{Left-Right Sfermion Mixing}
\label{sec:LR}

Although we have assumed negligible left-right sfermion mixing, there
may in general be some mixing, and this will alter both $\Neut \Neut$
annihilation through the process of \figref{annihilationQQLR} and
$\sigmaSI$ through the process of \figref{scatteringQLR}.  Even if
left-right mixing is only significant for third-generation sfermions,
as is typically true, in principle this may affect $\sigmaSI$ by
modifying $\Neut \Neut$ annihilation in the early Universe or by
changing the heavy quark contributions to scattering off nucleons.

\begin{figure}[tbp]
  \includegraphics[width=.48\textwidth]{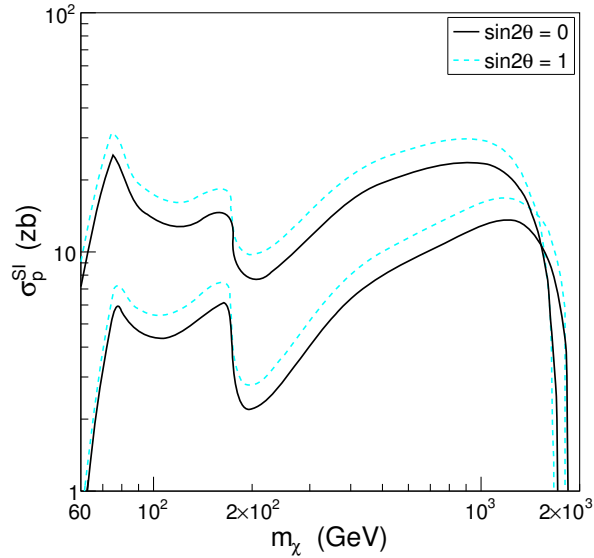}
\caption{{\em Dependence of $\sigmaSI$ on left-right sfermion mixing.}
  Cross sections $\sigmaSI$ are given for no left-right mixing and
  maximal left-right mixing for 3rd generation sfermions, $\mu >0$
  (upper) and $\mu<0$ (lower), $\mtilde = 2~\tev$, $m_A = 4~\tev$, and
  $\tan\beta = 10$.}
\label{fig:LR}
\end{figure}

\Figref{LR} shows the effect on $\sigmaSI$ of left-right mixing in
third generation sfermions.  The mixing here is parameterized by the
angle $\theta_{\tilde{q}}$, and the two cases shown are those with no
mixing ($\sin 2\theta_{\tilde{q}} = 0$) and maximal mixing ($\sin
2\theta_{\tilde{q}} = 1$), for which each mass eigenstate has equal
parts $\tilde{f}_L$ and $\tilde{f}_R$.  These are achieved by fixing
the $A_{t, b, \tau}$ parameters appropriately.  Although the latter
case does produce some splitting between the $\tilde{f}$ mass
eigenstates, for the degenerate masses considered here, a highly mixed
scenario can be achieved without a kinematically significant splitting
for our analysis.\footnote{The masses are not exactly degenerate,
given corrections from $D$-terms, but even including these, a pair of
completely mixed states can be created with a mass splitting of $\alt
2 - 3~\gev$ for the values of $\mtilde$ considered.  These corrections
are kept even in the no mixing case to prevent unintentional
left-right mixing from numerical effects.}

As seen in \figref{LR}, for $\mtilde = 2~\tev$, left-right mixing may
enhance $\sigmaSI$ by $\sim 15\%$.  The effects of left-right mixing
on the relic density calculation are minimal, but contributions from
$\Bino q_{L,R} \to \Bino q_{R,L}$ and $\Higgsino q_{L,R} \to \Higgsino
q_{R,L}$ produce an enhancement in $\sigmaSI$ directly.  The effect
may be larger for lower $\mtilde$, but in any case, we find that the
overall effect does not significantly impact conclusions about the
characteristic range of $\sigmaSI$.  The value of $\sigmaSI$ is
enhanced as $\chi$ becomes pure $\Higgsino$ at high mass by a process
similar to \figref{scatteringQLR} with $\Higgsino$ initial and final
states, but this process is yukawa suppressed and $\sigmaSI$ still
drops below $1~\zb$ in the highly $\Higgsino$ region.

\subsection{Explicit Inclusion of CP-Violation}
\label{sec:cpinclusion}

Although the desire for ${\cal O}(1)$ CP-violating phases motivates
large sfermion masses, we have performed our analysis is a manifestly
CP-conserving framework.  This is a cause for some concern, as CP
violation could impact relic density and scattering calculations,
which could, in principle, suppress $\sigmaSI$.  In our model
framework all CP violation is found in the phases of the parameters
$M_1$ and $\mu$.  We adopt a convention where all CP violation is
contained in the phase of the $\mu$ parameter $\phi_\mu$.

The effects of general phases on both $\Omega$ and $\sigmaSI$ were
discussed by the authors of Ref.~\cite{Nihei:2004bc}, who found that
phases can significantly enhance the relic density and suppress
$\sigmaSI$.  (See also Ref.~\cite{Belanger:2006qa}.) For a highly-mixed
neutralino LSP like the case considered here, varying the phase causes
a suppression in both annihilation and scattering cross sections,
leading to an enhancement in the relic density by a factor of $\sim 5$
and a suppression in $\sigmaSI$ by a factor of $\sim 50$.  However, in
that analysis the $\Omega_\chi$ and $\sigmaSI$ were treated as two
independent parameters.  In our analysis, with an appropriate relic
density applied as a constraint, these effects tend to cancel each
other: to achieve a viable relic density, $\ah$ must be increased to
enhance annihilation, which will in turn enhance $\sigmaSI$.

To determine the effect, we note that the relic density is nearly
minimal for $\phi_\mu = 0$ and maximal for $\phi_\mu = \pi$, with
intermediate values for $\phi_\mu$ smoothly interpolating between
these two limits.  The value of $\sigmaSI$ has opposite behavior, with
a maximal value for $\phi_\mu = 0$ and minimal value for $\phi_\mu =
\pi$.  The two limiting cases are, then, precisely the cases $\mu > 0$
and $\mu < 0$ that we have considered throughout our analysis.  Thus
the inclusion of an arbitrary CP-violating parameter should result in
a simple interpolation between the cases of $\mu > 0$ and $\mu < 0$ as
alluded to in \secref{main}, without resulting in values of $\sigmaSI$
outside of this range, except perhaps in finely-tuned portions of
parameter space.

\subsection{Strange Quark Content of the Nucleon}
\label{sec:fs}

As discussed in \secref{dd}, the value of $\sigmaSI$ is strongly
dependent on the individual quark contributions to the nucleon,
$f_q^N$.  Of these, the quantity that has traditionally had the most
uncertainty is $f_s^N$, with quoted values ranging from $0.05$ to
$0.23$.  Recent lattice results suggest the lower value, and we have
used this value throughout, but it is useful to examine the variation
of $\sigmaSI$ as $f_s^N$ is varied.  Throughout this section we use
the assumption $f_q^p \approx f_q^n$.

\begin{figure}[tbp]
  \subfigure[$\mtilde = 2~\tev$]{
    \includegraphics[width=.48\textwidth]{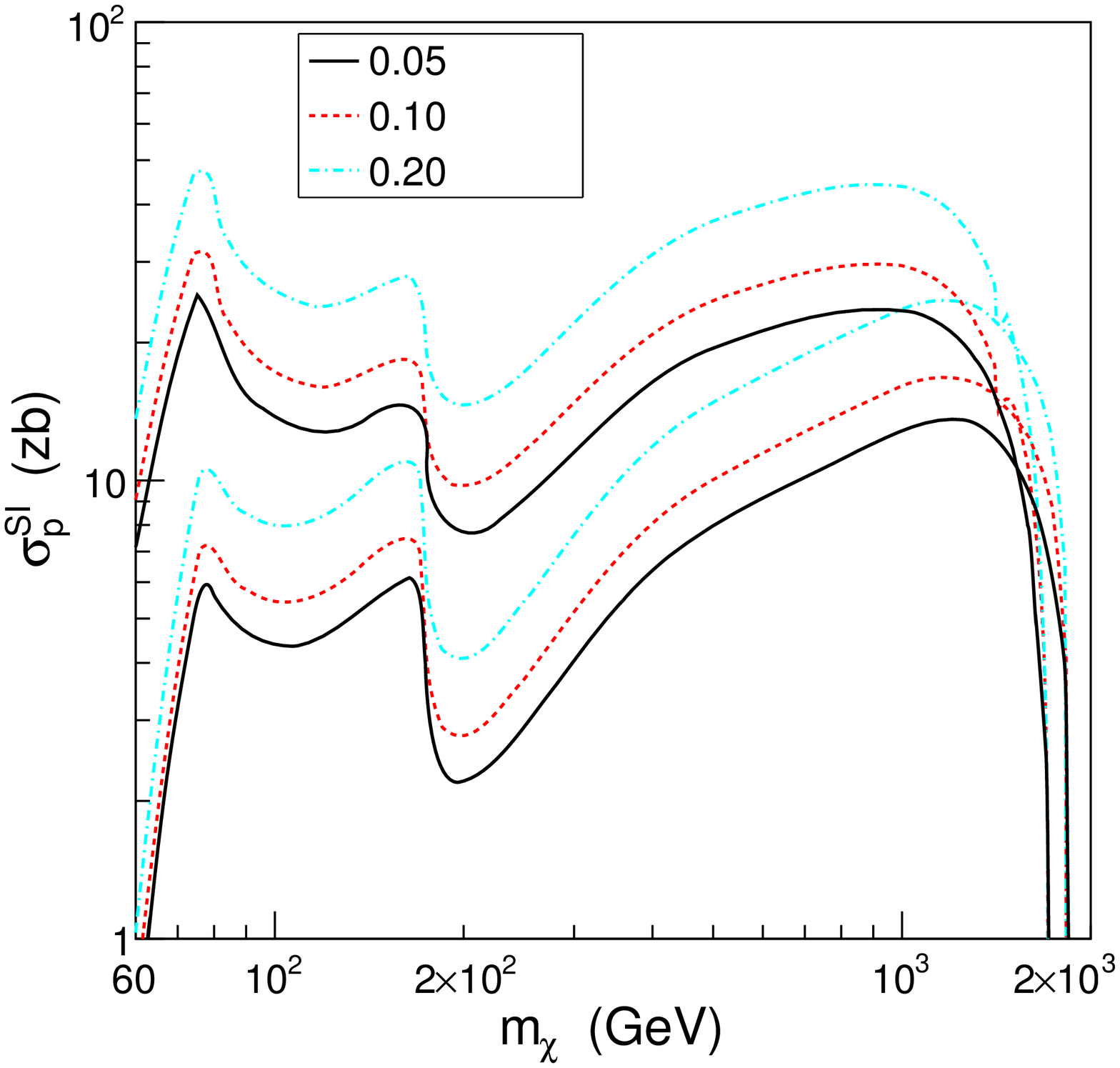}
    \label{fig:strange2tev} }
  \subfigure[$\mtilde = 300~\gev$]{
    \includegraphics[width=.48\textwidth]{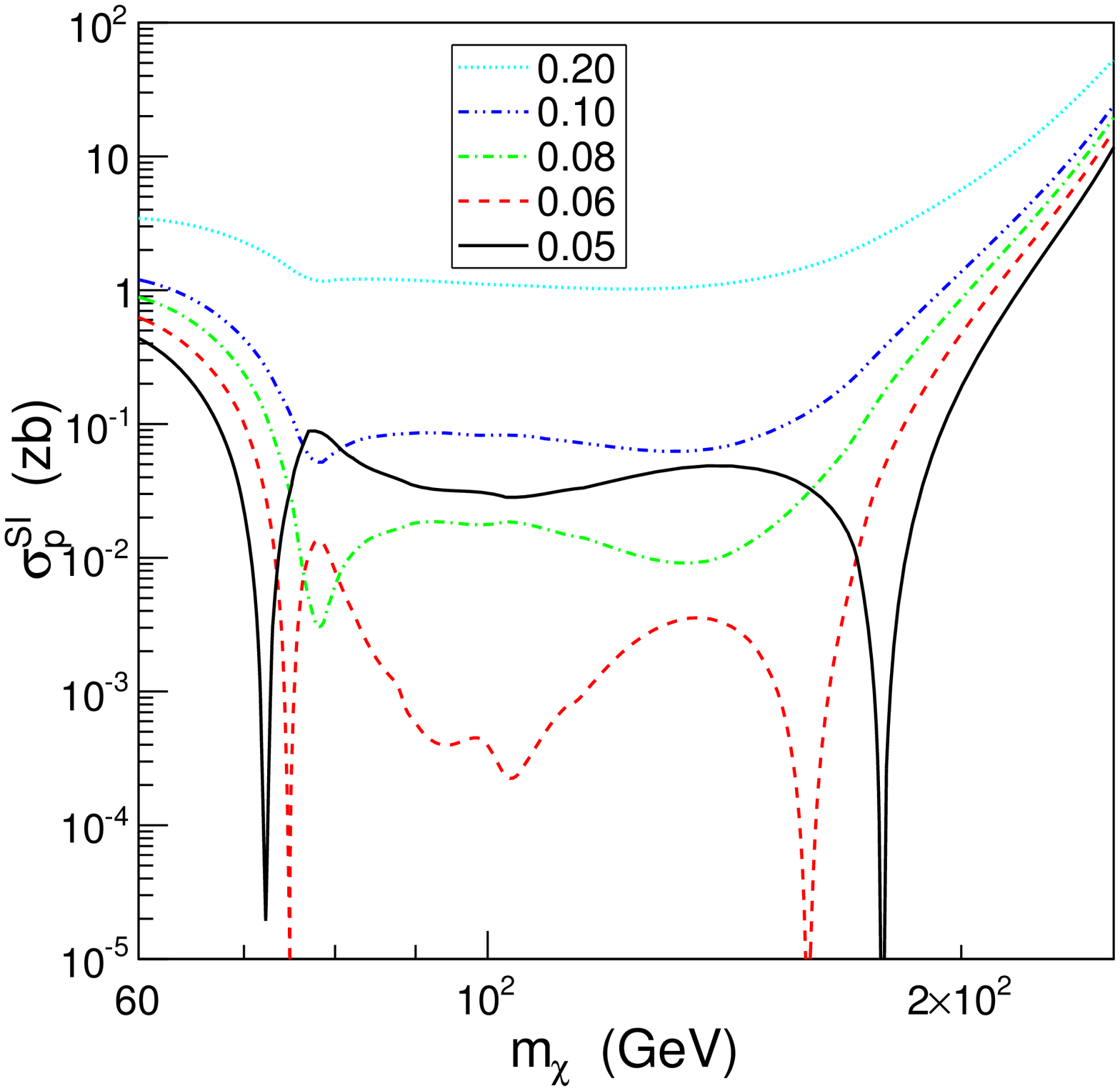}
    \label{fig:strange300gev} }
\caption{{\em Dependence of $\sigmaSI$ on the strange quark content of
the nucleon.} Cross sections are given for the different values of
$f_s^p$ indicated, $m_A = 4~\tev$, and $\tan\beta = 10$.  In panel
\subref{fig:strange2tev}, $\mtilde = 2~\tev$ and $\mu > 0$ (upper) and
$\mu < 0$ (lower).  In panel \subref{fig:strange300gev}, $\mtilde =
300~\gev$, and only the $\mu < 0$ curves are shown, as the $\mu > 0$
curves have much the same behavior as in panel
\subref{fig:strange2tev}.}
\label{fig:strange}
\end{figure}

To understand the general dependence on $f_s^N$, consider the quantity
\begin{equation}
f_{\text{tot}}^N = \sum\limits_{q=1}^6 f_q^N \ ,
\end{equation}
which can be interpreted as the total quark contribution to the
nucleon mass.  The (perhaps unreasonably) high value $f_s^N = 0.23$
implies $f_{\text{tot}}^N \simeq 0.44$, and the low value $f_s^N =
0.05$ implies $f_{\text{tot}}^N \simeq 0.30$.  Since $f_Q^N$ is
largely unchanged as $f_s^N$ decreases further, even the unrealistic
case of $f_s^N = 0$ would still have $f_{\text{tot}}^N \approx 0.25$.
Thus, without considering further possible cancellations, the
difference in $\sigmaSI$ between between no strange quark contribution
to the nucleon and the largest proposed value should be about
$(0.44/0.25)^2 \approx 3$.  A more careful variation is shown in
\figref{strange} for $\mtilde = 2~\tev$ and $300~\gev$.  The $2~\tev$
case matches the sketched analysis above quite well --- the value of
$\sigmaSI$ for $f_s^N = 0.05$ is suppressed by $\sim 2$ relative to
the value for $f_s^N = 0.2$, with the value for $f_s^N = 0.1$ lying in
between, in good agreement with the prediction based on
$f_{\text{tot}}^N$.  Note that, for $\mtilde = 2~\tev$, our default
value of $f_s^N$ yields the lowest $\sigmaSI$, and other possible
values imply improved prospects for direct detection.

For $\mtilde = 300~\gev$ the suppression for low $f_s^N$ is much more
pronounced, at least in the case of $\mu < 0$.  As discussed in
\secref{dd}, for $\mu < 0$ the sfermion contribution to scattering is
similar in size to the Higgs contribution, opening the possibility of
cancellation regardless of the value of $\mneut$.  In fact, this also
opens up the possibility for additional cancellations between
$\lambda_q$ for different quarks based on relative values of $f_q^p$.
The cancellation is most prominent when $f_s^N$ is comparable to the
contributions of other quarks; for larger values of $f_s^N$ the
strange quark is the dominant contribution to nucleon mass, which
raises the generic value of $\sigmaSI$.  Even for larger $f_s^N$,
however, $\sigmaSI$ for $\mtilde = 300~\gev$ remains suppressed
relative to higher values of $\mtilde$ due to the squark-Higgs
contribution cancellations discussed in \secref{dd}.

\subsection{Variations in Relic Density}
\label{sec:astro}

It is possible that neutralinos form some, but not all, of the dark
matter. One may consider models with $\Omegachi < 0.23$, which may
have larger $\sigmaSI$.  Of course the direct detection signal is
proportional not to $\sigmaSI$, but to $\Omegachi \sigmaSI$.  In fact,
as is well-known, for thermal relics, the relic density is inversely
proportional to $\langle \sigmaan v \rangle$, the thermally-averaged
annihilation cross section.  If $\sigmaSI$ and $\sigmaan$ scale
together, then the gain in $\sigmaSI$ is exactly compensated by the
reduction in $\Omegachi$.

\begin{figure}[tbp]
  \subfigure[$\sigmaSI$]{
    \includegraphics[width=.48\textwidth]{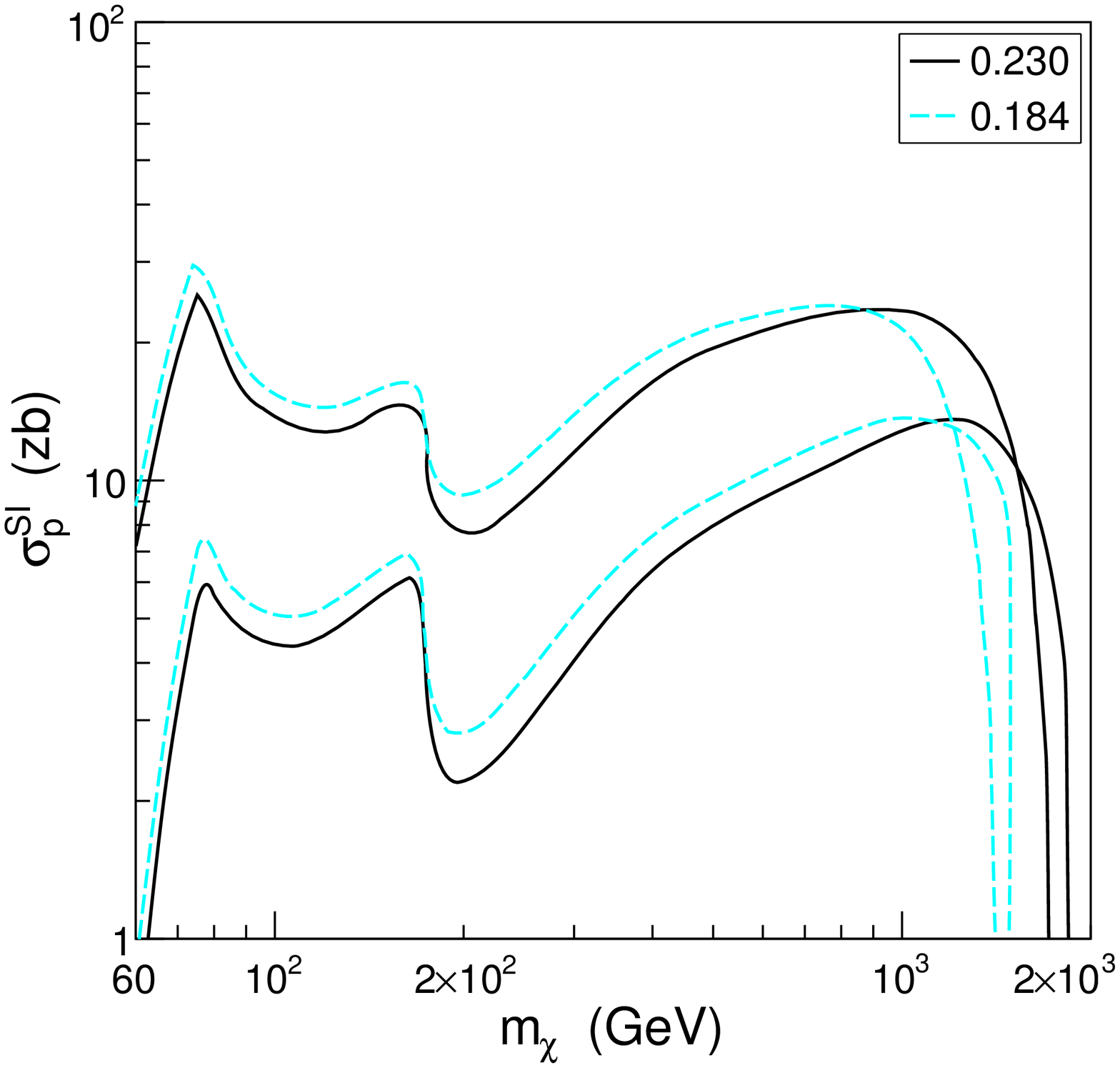}
    \label{fig:omegaonly} }
  \subfigure[$\Omegachi \sigmaSI$]{
    \includegraphics[width=.48\textwidth]{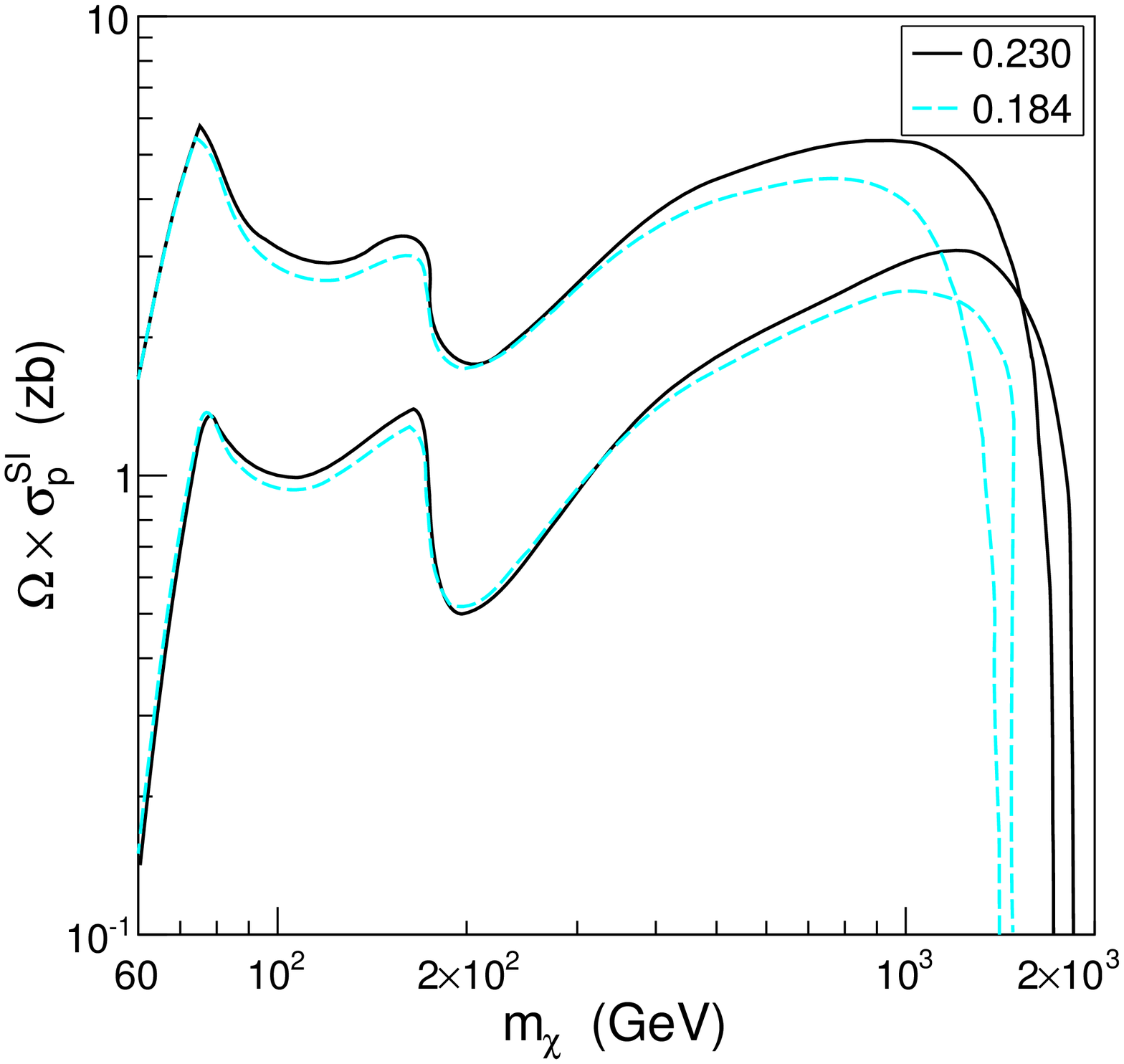}
    \label{fig:omegasigma} }
\caption{{\em Dependence of $\sigmaSI$ on $\Omegachi$.}  Cross
sections $\sigmaSI$ are given for the values of $\Omegachi$ indicated,
$\mu >0$ (upper) and $\mu<0$ (lower), $\mtilde = 2~\tev$, $m_A =
4~\tev$, and $\tan\beta = 10$.}
\label{fig:omega}
\end{figure}

In \figref{omega}, we explore the impact of variations in $\Omegachi$.
For concreteness, we consider models with $\Omegachi = 0.184$.  As
expected, in \figref{omegaonly}, we see that these models have larger
$\sigmaSI$.  However, in \figref{omegasigma}, we see that this is
almost exactly compensated by the reduction in $\Omegachi$ except at
high $m_\chi$, leaving the direct detection signal invariant.  At very
high $m_\chi$, the neutralino becomes nearly pure $\Higgsino$ and
reaches its maximum annihilation efficiency more quickly for the lower
value of $\Omega$.  For relatively small deviations from the
assumption $\Omegachi = 0.23$, then, the direct detection signal is
not impacted by the exact value of $\Omegachi$ assumed except for
large $m_\chi$.  This result may not hold for large deviations when
$\Neut$ constitutes a very small portion of the dark matter, but in
this case the theoretical motivation for $\Neut$ dark matter is
somewhat strained.

\subsection{Galactic Small-Scale Structure}

The exclusion limits on $\sigmaSI$ are also affected by the local
density of DM.  Although a na\"{i}ve approximation suggests a
spherical DM halo with uniform density at a given radius, small scale
structure exists, and leads to a few regions with higher than average
density, compensated for by a lower than average density in the rest
of the space.  This has been studied in
Ref.~\cite{Kamionkowski:2008vw}, which concluded that while the
reduction in general number density from clumping might be as much as
an order of magnitude, the more likely possibility is a reduction by
less than a factor of two.  This implies that experimental limits are
subject to astrophysical uncertainties, but does not significantly
alter our conclusion that near future experiments will probe the heart
of neutralino parameter space.

\subsection{Extension to Non-minimal Scenarios}
\label{sec:extension}

Because we work in a manifestly low-energy framework, our results can
easily be generalized beyond the MSSM to models that include new
superfields with $\tev$-scale masses.  Although such fields will in
general affects the annihilation and scattering of $\Neut$ dark
matter, we argue that $\sigmaSI$ remains generically in the range
$1~\zb \alt \sigmaSI \alt 40~\zb$.

To illustrate this point, consider first the impact of a new
\tev-scale particle $X$ on annihilation.  Assume that $X$ couples to
$\Neut$ and some set of other MSSM particles with a coupling of order
$\alpha_W$.  The effect of $X$ will be most pronounced if there is
resonant annihilation when $m_X \approx 2 \mneut$ or if there is
significant co-annihilation when $m_X \approx \mneut$.  Both cases
force a severe suppression in $\ah$ to compensate and produce a
sufficiently large relic density\footnote{If reducing $\ah$ does not
compensate for the resonant annihilation, then that region will
generally not provide a viable relic density.}, in turn causing a
large suppression in $\sigmaSI$ similar to that seen in the $h$ and
$A$ resonances discussed above.  However, outside of these fine-tuned
regions of parameter space, new $X$-mediated processes must be roughly
the same size as existing processes to be relevant and will generally
decrease (or possibly increase in the case of cancellation) $\ah$
relative to the MSSM case.  Although this will produce a suppression
(or possibly enhancement) of $\sigmaSI$, the additional effect should
not generically be greater than the factor of $2 - 3$ that is caused
by the introduction of $W^+ W^-$ or $t \bar{t}$ annihilation.

In scattering, if $X$ couples to quarks, it may induce additional
contributions to $\sigmaSI$.  However, although the mass and
interactions of $X$ may be chosen to significantly suppress
$\sigmaSI$, such an effect requires a fine-tuning of the mass and
couplings of $X$ to produce a cancellation similar to the case of
$\mtilde = 300~\gev$ discussed in \secref{dd}.  Indeed, the low value
of $m_X$ required to cause a severe suppression directly through
scattering effects should na\"{i}vely interfere with low-energy
observables due to its couplings to quarks, and such operators are
constrained by collider
searches~\cite{Goodman:2010yf,Goodman:2010ku,Bai:2010hh}.  Even if
constraints on the mass of $X$ can be avoided by tuning couplings, a
sufficiently light mass for $X$ places it generically within detection
reach of the LHC.

\section{Spin-Dependent Cross Sections}
\label{sec:sd}

We have focused in this work on spin-independent direct detection, as
this provides bright prospects for the discovery of neutralino dark
matter.  It is interesting, nonetheless, to examine the prospects for
spin-dependent cross sections, which is probed both by direct
detection experiments, and by indirect detection searches for
neutrinos.

Spin-dependent scattering proceeds through the effective operator
\begin{equation}
\alpha_i \bar{\Neut} \gamma^\mu \gamma^5 \Neut \bar{q}_i \gamma_\mu
\gamma^5 q_i\ .
\end{equation}
The structure of these operators implies that the contributing
diagrams will be similar to those shown in \figref{scattering}, except
with $Z$-mediated scattering replacing Higgs-mediated scattering, and
the replacement $\Bino \leftrightarrow \Higgsino$ and associated
fermion $L \leftrightarrow R$ replacements on one leg in both
\figref{scatteringQ} and \figref{scatteringQLR}.  The associated
coefficients have the form~\cite{Ellis:2000jd}
\begin{eqnarray}
\nonumber \alpha_u &=& \frac{g^2}{2 \left(m_{\tilde{u}_i}^2 -
  \mneut^2 \right)} \left[ \left| \left( \frac{1}{2} \aw + \frac{1}{6}
\tan\theta_W \ab\right) U^{(\tilde{u})*}_{i1} + \frac{m_u \ahu}{2 m_W
  \sin\beta} U^{(\tilde{u})*}_{i2} \right|^2 \right. \\
\nonumber & & \hspace*{5cm} + \left. \left| \frac{m_u \ahu}{2 m_W
\sin\beta} U^{(\tilde{u})}_{i1} - \frac{2}{3} \tan\theta_W \ab
U^{(\tilde{u})}_{i2} \right|^2 \right] \\
& & - \frac{g^2}{8 m_Z^2 \cos^2 \theta_W} \left(
\left|\ahd\right|^2 - \left|\ahu\right|^2 \right) \\
\nonumber \alpha_d &=& \frac{g^2}{2 \left(m_{\tilde{d}_i}^2 -
  \mneut^2 \right)} \left[ \left| \left( - \frac{1}{2} \aw + \frac{1}{6}
  \tan\theta_W \ab \right) U^{(\tilde{d})*}_{i1} + \frac{m_d \ahd}{2
  m_W \cos\beta} U^{(\tilde{d})*}_{i2} \right|^2 \right. \\
\nonumber & & \hspace*{5cm} + \left. \left| \frac{m_d \ahd}{2 m_W
\cos\beta} U^{(\tilde{d})}_{i1} + \frac{1}{3} \tan\theta_W \ab
U^{(\tilde{d})}_{i2} \right|^2 \right] \\
& & + \frac{g^2}{8 m_Z^2 \cos^2 \theta_W} \left( \left|\ahd\right|^2 -
\left|\ahu\right|^2 \right) \ ,
\end{eqnarray}
with the same notation used in \secref{lambdaq}.

The spin-dependent couplings will exhibit a similar suppression of
sfermion-mediated diagrams relative to $Z$-mediated diagrams as seen
in the spin-independent case, with the smaller value of $m_Z$ relative
to $m_h$ leading to the na\"{i}ve expectation that the squark mass
necessary to cause serious suppression should be lower than the
$300~\gev$ mass necessary to suppress $\sigmaSI$.  Such a small mass
is ruled out by collider bounds.  Moreover, as long as $\ahud$ are
non-negligible, the squark-mediated contribution to the cross section
suffers an additional suppression relative to the $Z$-mediated
contribution of
\begin{equation}
 \sim \sin^4 \theta_W \left(\frac{\left| \ab
   \right|^2}{\left|\ahd\right|^2 - \left|\ahu\right|^2}\right)^2\ ,
\end{equation}
in contrast to the minor suppression of the Higgs-mediated
contribution relative to the squark-mediated contribution due to the
mixing angle $\alpha$ for $\sigmaSI$.

\begin{figure}[tbp]
  \includegraphics[width=.72\textwidth]{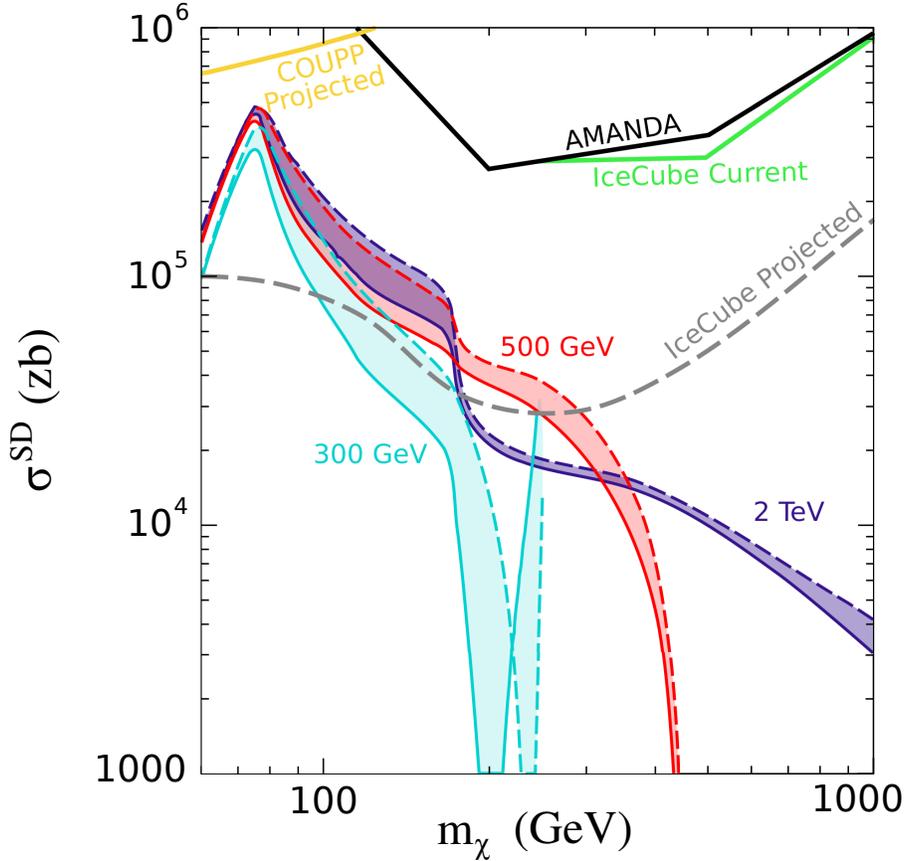}
\caption{{\em Characteristic values for spin-dependent
neutralino-proton scattering cross sections $\sigmaSD$.}  The shaded
regions are $\sigmaSD$ for the values $\mtilde$ indicated, with $\mu >
0$ (solid border) and $\mu < 0$ (dashed border), $m_A = 4~\tev$, and
$\tan\beta = 10$. Also shown are current limits from the
AMANDA~\cite{Achterberg:2006jf} and IceCube~\cite{Abbasi:2009uz}
experiments, and the projected sensitivities of a 10-year run of
IceCube with DeepCore~\cite{Braun:2009fr} and a 4kg COUPP running at a
deep underground site~\cite{Behnke:2010xt}.}
\label{fig:sd}
\end{figure}

Predictions for $\sigmaSD$ for neutralino-proton scattering are given
in \figref{sd}, using the same parameters as in \secref{main} and
three values of $\mtilde$.  In all three cases, the dominant effect is
a drop in $\sigmaSD$ with increasing $\mneut$.  This is due to the
proportionality of the $Z$-mediated contribution to the factor
$|\ahd|^2 - |\ahu|^2$; as demonstrated in \figref{ah2TeV}, the values
of $|\ahud|$ converge with increasing $\mneut$, thus suppressing
$\sigmaSD$.  The dependence on $\mtilde$ is mild until higher values
of $\mneut$ where the $Z$-mediated contribution is self-suppressed to
the level of the sfermion-mediated contribution, and even the
differences in $\sigmaSD$ at $\mneut < 100~\gev$ are due to the slight
variation in $\ah$ with mass shown in \figref{ahvarious} rather than
cancellations between scattering processes.

The limits plotted in \figref{sd} correspond to upper limits on
$\sigmaSD$ from the AMANDA~\cite{Achterberg:2006jf} and
IceCube~\cite{Abbasi:2009uz} experiments, along with projected upper
limits from a 10 year run of IceCube with DeepCore, an extension of
IceCube with tighter string and optical module
spacing~\cite{Braun:2009fr} and a 4 kg COUPP, assumed to run
background free for 3 months at a deep underground
site~\cite{Behnke:2010xt}.  The entire characteristic region plotted
lies outside current upper limits, but the projected IceCube limits
intersect the allowed region for $60~\gev \alt \mneut \alt 200~\gev$
for most values of $\mtilde$, and the $\mtilde=500~\gev$ region may be
within IceCube's sensitivity region up to $\mneut \sim 300~\gev$.

We conclude that, at least for the neutralino DM models considered
here, spin-dependent searches are indeed not as promising as
spin-independent searches in the near future.  However, for some parts
of parameter space, spin-dependent searches may also see a signal in
the coming years.  Of course, if a signal is seen in spin-independent
detection, confirmation through spin-dependent signals will be an
essential supplement.

\section{Conclusions}
\label{sec:conclusion}

Experimental sensitivities to the scattering of dark matter off normal
matter are improving rapidly.  In this study, we have addressed the
question of what this progress implies for the discovery prospects for
neutralino dark matter.  To determine characteristic spin-independent
and spin-dependent neutralino-nucleon cross sections, we have worked
not with specific high-energy models, but rather in a phenomenological
framework incorporating only the basic motivations for SUSY (the gauge
hierarchy problem, force unification, and the WIMP miracle) and
experimental constraints (direct searches and bounds on flavor and CP
violation).

SUSY parameter space is large and complicated, but our strategy has
been to start in a simple corner and work out from there.  The
conclusions of this study may be summarized as follows.  We assume
gaugino mass unification, as motivated by gauge coupling unification,
and, to begin with, also heavy squarks and sleptons, as motivated by
stringent constraints on flavor and CP violation.  All of neutralino
dark matter phenomenology then depends only on the Bino and Higgsino
mass parameters $M_1$ and $\mu$, and the heavy Higgs scalar mass $m_A$
and $\tan\beta$.  Away from Higgs resonances, the dependence on the
latter two parameters is weak.  Requiring the correct thermal relic
density removes another degree of freedom, leaving only one, which may
be taken to be the neutralino mass $\mneut$.  For $\mneut \agt
70~\gev$, the spin-independent cross section lies in the range $1~\zb
\alt \sigmaSI \alt 40~\zb$ and is highly insensitive to $\mneut$ for
the reasons detailed above.  The assumption of a unified sfermion mass
disregards the complicated spectrum details associated with any
high-scale model, but according to the discussion above such effects
should be sub-dominant in any case except in very special cases such
as co-annihilation.  While the mass splittings of full spectra would
certainly alter a specific prediction for $\sigmaSI$, such effects
will not alter the {\it characteristic} value of $\sigmaSI$ studied
here.

From this result, we then generalize to other regions of parameter
space.  Surprisingly, the upper and lower limits of the characteristic
range only vary by factors of $\sim 2$ for sfermion masses as low as
$\mtilde \sim 400~\gev$.  For very light sfermions near their
experimental bound $\mtilde \sim 300~\gev$, large cancellations
between contributions from up- and down-type quarks are possible.  We
have also identified and analyzed two other fine-tuned scenarios with
greatly reduced $\sigmaSI$, where annihilation is enhanced either by
resonances when $2 \mneut \approx m_h$ or $m_A$ or by co-annihilation
when $\mneut \approx \mtilde$. In these special cases, $\sigmaSI$
falls outside our characteristic range.

We next considered the effects of un-unifying the slepton and squark
masses, un-unifying the third generation squark masses, including
significant left-right sfermion mixing, and including CP-violating
phases in the relic density and $\sigmaSI$ calculations.  In each of
these cases, for fixed $\Omegachi$, we find typically small variations
in the allowed range of $\sigmaSI$.  We have also explored variations
of the strange quark content of the nucleon and found that our assumed
value of $f_s^N = 0.05$ is conservative, with other values enhancing
$\sigmaSI$ for $\mtilde \agt 400~\gev$. Finally, we also considered
the possibility of multi-component dark matter where neutralinos are
just one significant component.  In this case, neutralino annihilation
must be enhanced to reduce the thermal relic density, but this also
enhances $\sigmaSI$.  These effects effectively cancel, leaving the
experimental direct detection signal, which is proportional to
$\Omegachi \sigmaSI$, invariant and in our characteristic range.

These results for $\sigmaSI$ are compared to experiment in
\figref{sigmas_limits}.  If dark matter is composed of thermal relic
neutralinos, the prospects for near future experiments that are
sensitive to the zeptobarn scale are truly promising.  We have also
analyzed the prospects for spin-dependent scattering.  These are
summarized in \figref{sd}.  Although not as promising as
spin-independent scattering, observable signals exist for dark matter
masses $\mneut \alt 500~\gev$.

Our results are complementary to those based on high-energy
frameworks.  In the context of mSUGRA, for example, the fine-tuned
possibilities of resonant annihilation and co-annihilation also exist
and are known to predict low $\sigmaSI$.  This results of this study
show that these phenomena are decoupled from other SUSY phenomenology
typically related to them in mSUGRA.  For example, the $A$ resonance
may be realized at any $\tan\beta$, as opposed to only high
$\tan\beta$ in mSUGRA.  Nevertheless, these possibilities remain
fine-tuned, in the sense of requiring specific relations between
superpartner masses that are not suggested by experimental constraints
or foundational motivations of SUSY.

At the same time, this study shows that results from the focus point
region of mSUGRA are robust.  As is well-known, in the focus point
region, the neutralino is a well-mixed Bino-Higgsino combination and,
for this reason, rates for both direct and indirect detection are
large, with predictions for spin-independent cross sections in the
range $\sigmaSI \sim 1 - 10~\zb$.  The results of this study show that
similar cross sections are achieved in a more general setting
throughout parameter space, largely independent of assumptions about
GUT boundary conditions, $A$ parameters, $\tan\beta$, or even the
details of the slepton and squark spectra, provided $\mtilde \agt
400~\gev$.  Large direct detection signals and light sfermions are
therefore perfectly consistent, with positive implications for both
dark matter experiments and colliders.  Given the generality of our
model framework, the results for characteristic $\sigmaSI$ will apply
to a diverse array of high-energy models.

In conclusion, the field is reaching a critical juncture in the search
for neutralino dark matter.  We have identified a range for
characteristic cross sections for neutralino-nucleon scattering.
These are by no means a lower limit on possible value of $\sigmaSI$;
indeed several possibilities for much smaller cross-sections were
identified in this work.  Nevertheless, even in the most pessimistic
scenario, the identification of a characteristic cross section range
has important consequences.  If no signal is seen at the zeptobarn
scale, then a model with thermal relic neutralinos must contain some
amount of fine-tuning.  This could be accomplished by enhanced
annihilation through resonances or co-annihilation, as noted above.
Alternatively, one could introduce new particles that induce
neutralino-quark couplings that cancel existing contributions to
$\sigmaSI$.  As we have seen, however, the scale of existing
contributions is set by the light Higgs boson mass $m_h$, not by a
superpartner mass.  These new contributions would typically be
associated with light, colored fields, then, as in the case of the
$\sim 300~\gev$ squarks we discussed in detail, and would be within
reach of collider experiments.

Of course, the most straightforward and natural conclusion is that if
thermal relic neutralinos are a significant component of dark matter,
their spin-independent scattering cross section likely lies in the
range $1~\zb \alt \sigmaSI \alt 40~\zb$, just beyond current
sensitivities.  Current and near future experiments will therefore see
a signal as they probe down to the zeptobarn scale, and the next
generation of experiments will compile the statistics required to
definitively establish the signal on different targets, constrain the
dark matter's mass, see signals in directional detectors, and usher in
the era of dark matter astronomy.

\section*{Acknowledgments}

We are grateful to Jeff Filippini, Konstantin Matchev, Will Shepherd,
Hank Sobel, Tim Tait, and Xerxes Tata for helpful correspondence and
conversations.  This work was supported in part by NSF grants
PHY--0653656 and PHY--0970173.


\bibliography{dmzb05}{}

\providecommand{\href}[2]{#2}\begingroup\raggedright\begin{thebibliography}{10%
0}

\bibitem{Bertone:2004pz}
G.~Bertone, D.~Hooper, and J.~Silk, ``{Particle dark matter: Evidence,
  candidates and constraints},''
  \href{http://dx.doi.org/10.1016/j.physrep.2004.08.031}{{\em Phys. Rept.}
  {\bfseries 405} (2005) 279--390},
\href{http://arxiv.org/abs/hep-ph/0404175}{{\ttfamily arXiv:hep-ph/0404175}}.

\bibitem{Bergstrom:2009ib}
L.~Bergstrom, ``{Dark Matter Candidates},''
  \href{http://dx.doi.org/10.1088/1367-2630/11/10/105006}{{\em New J. Phys.}
  {\bfseries 11} (2009) 105006},
\href{http://arxiv.org/abs/0903.4849}{{\ttfamily arXiv:0903.4849 [hep-ph]}}.

\bibitem{Feng:2010gw}
J.~L. Feng, ``{Dark Matter Candidates from Particle Physics and Methods of
  Detection},''
  \href{http://dx.doi.org/10.1146/annurev-astro-082708-101659}{{\em Ann. Rev.
  Astron. Astrophys.} {\bfseries 48} (2010) 495--545},
\href{http://arxiv.org/abs/1003.0904}{{\ttfamily arXiv:1003.0904
  [astro-ph.CO]}}.

\bibitem{Bernabei:2010mq}
R.~Bernabei {\em et al.}, ``{New results from DAMA/LIBRA},''
  \href{http://dx.doi.org/10.1140/epjc/s10052-010-1303-9}{{\em Eur. Phys. J.}
  {\bfseries C67} (2010) 39--49},
\href{http://arxiv.org/abs/1002.1028}{{\ttfamily arXiv:1002.1028
  [astro-ph.GA]}}.

\bibitem{Angle:2008we}
{\bfseries XENON10} Collaboration, J.~Angle {\em et al.}, ``{Limits on
  spin-dependent WIMP-nucleon cross-sections from the XENON10 experiment},''
  \href{http://dx.doi.org/10.1103/PhysRevLett.101.091301}{{\em Phys. Rev.
  Lett.} {\bfseries 101} (2008) 091301},
\href{http://arxiv.org/abs/0805.2939}{{\ttfamily arXiv:0805.2939 [astro-ph]}}.

\bibitem{Ahmed:2009zw}
{\bfseries The CDMS-II} Collaboration, Z.~Ahmed {\em et al.}, ``{Dark Matter
  Search Results from the CDMS II Experiment},''
  \href{http://dx.doi.org/10.1126/science.1186112}{{\em Science} {\bfseries
  327} (2010) 1619--1621},
\href{http://arxiv.org/abs/0912.3592}{{\ttfamily arXiv:0912.3592
  [astro-ph.CO]}}.

\bibitem{Aalseth:2010vx}
{\bfseries CoGeNT} Collaboration, C.~E. Aalseth {\em et al.}, ``{Results from a
  Search for Light-Mass Dark Matter with a P- type Point Contact Germanium
  Detector},''
\href{http://arxiv.org/abs/1002.4703}{{\ttfamily arXiv:1002.4703
  [astro-ph.CO]}}.

\bibitem{Aprile:2010um}
{\bfseries XENON100} Collaboration, E.~Aprile {\em et al.}, ``{First Dark
  Matter Results from the XENON100 Experiment},''
\href{http://arxiv.org/abs/1005.0380}{{\ttfamily arXiv:1005.0380
  [astro-ph.CO]}}.

\bibitem{Aprile:2011hx}
{\bfseries XENON100} Collaboration, E.~Aprile {\em et al.}, ``{Likelihood
  Approach to the First Dark Matter Results from XENON100},''
\href{http://arxiv.org/abs/1103.0303}{{\ttfamily arXiv:1103.0303 [hep-ex]}}.

\bibitem{Aprile:2011hi}
{\bfseries XENON100} Collaboration, E.~Aprile {\em et al.}, ``{Dark Matter
  Results from 100 Live Days of XENON100 Data},''
\href{http://arxiv.org/abs/1104.2549}{{\ttfamily arXiv:1104.2549
  [astro-ph.CO]}}.

\bibitem{Aprile:2009yh}
{\bfseries XENON100} Collaboration, E.~Aprile, L.~Baudis, {\em et al.},
  ``{Status and Sensitivity Projections for the XENON100 Dark Matter
  Experiment},'' {\em PoS} {\bfseries IDM2008} (2008) 018,
\href{http://arxiv.org/abs/0902.4253}{{\ttfamily arXiv:0902.4253
  [astro-ph.IM]}}.

\bibitem{Schmaler:2009bu}
{\bfseries CRESST} Collaboration, J.~Schmaler {\em et al.}, ``{Status of the
  CRESST Dark Matter Search},''
\href{http://arxiv.org/abs/0912.3689}{{\ttfamily arXiv:0912.3689
  [astro-ph.IM]}}.

\bibitem{McKinsey:2010zz}
{\bfseries LUX} Collaboration, D.~N. McKinsey {\em et al.}, ``{The LUX dark
  matter search},''
\href{http://dx.doi.org/10.1088/1742-6596/203/1/012026}{{\em J. Phys. Conf.
  Ser.} {\bfseries 203} (2010) 012026}.

\bibitem{Sekiya:2010bf}
{\bfseries XMASS} Collaboration, H.~Sekiya {\em et al.}, ``{XMASS},''
\href{http://arxiv.org/abs/1006.1473}{{\ttfamily arXiv:1006.1473
  [astro-ph.IM]}}.

\bibitem{Hime:2006zq}
{\bfseries DEAP/CLEAN} Collaboration, A.~Hime, ``{DEAP \& CLEAN detectors for
  low energy particle astrophysics},''
\href{http://dx.doi.org/10.1063/1.2402618}{{\em AIP Conf. Proc.} {\bfseries
  870} (2006) 205--207}.

\bibitem{Acciarri:2010zz}
{\bfseries WArP} Collaboration, R.~Acciarri {\em et al.}, ``{The WArP
  experiment},''
\href{http://dx.doi.org/10.1088/1742-6596/203/1/012006}{{\em J. Phys. Conf.
  Ser.} {\bfseries 203} (2010) 012006}.

\bibitem{Bruch:2010eq}
T.~Bruch and f.~t.~C. Collaboration, ``{CDMS-II to SuperCDMS: WIMP search at a
  zeptobarn},''
\href{http://arxiv.org/abs/1001.3037}{{\ttfamily arXiv:1001.3037
  [astro-ph.IM]}}.

\bibitem{Strigari:2009bq}
L.~E. Strigari, ``{Neutrino Coherent Scattering Rates at Direct Dark Matter
  Detectors},'' \href{http://dx.doi.org/10.1088/1367-2630/11/10/105011}{{\em
  New J. Phys.} {\bfseries 11} (2009) 105011},
\href{http://arxiv.org/abs/0903.3630}{{\ttfamily arXiv:0903.3630
  [astro-ph.CO]}}.

\bibitem{Gutlein:2010tq}
A.~Gutlein {\em et al.}, ``{Solar and Atmospheric Neutrinos: Background Sources
  for the Direct Dark Matter Searches},''
  \href{http://dx.doi.org/10.1016/j.astropartphys.2010.06.002}{{\em Astropart.
  Phys.} {\bfseries 34} (2010) 90--96},
\href{http://arxiv.org/abs/1003.5530}{{\ttfamily arXiv:1003.5530 [hep-ph]}}.

\bibitem{Goldberg:1983nd}
H.~Goldberg, ``{Constraint on the photino mass from cosmology},''
\href{http://dx.doi.org/10.1103/PhysRevLett.50.1419}{{\em Phys. Rev. Lett.}
  {\bfseries 50} (1983) 1419}.

\bibitem{Ellis:1983ew}
J.~R. Ellis, J.~S. Hagelin, D.~V. Nanopoulos, K.~A. Olive, and M.~Srednicki,
  ``{Supersymmetric relics from the big bang},''
\href{http://dx.doi.org/10.1016/0550-3213(84)90461-9}{{\em Nucl. Phys.}
  {\bfseries B238} (1984) 453--476}.

\bibitem{Drees:1992am}
M.~Drees and M.~M. Nojiri, ``{The Neutralino relic density in minimal $N=1$
  supergravity},'' \href{http://dx.doi.org/10.1103/PhysRevD.47.376}{{\em Phys.
  Rev.} {\bfseries D47} (1993) 376--408},
\href{http://arxiv.org/abs/hep-ph/9207234}{{\ttfamily arXiv:hep-ph/9207234}}.

\bibitem{Drees:1993bu}
M.~Drees and M.~Nojiri, ``{Neutralino-Nucleon Scattering Revisited},''
  \href{http://dx.doi.org/10.1103/PhysRevD.48.3483}{{\em Phys. Rev.} {\bfseries
  D48} (1993) 3483--3501},
\href{http://arxiv.org/abs/hep-ph/9307208}{{\ttfamily arXiv:hep-ph/9307208}}.

\bibitem{Jungman:1995df}
G.~Jungman, M.~Kamionkowski, and K.~Griest, ``{Supersymmetric dark matter},''
  \href{http://dx.doi.org/10.1016/0370-1573(95)00058-5}{{\em Phys. Rept.}
  {\bfseries 267} (1996) 195--373},
\href{http://arxiv.org/abs/hep-ph/9506380}{{\ttfamily arXiv:hep-ph/9506380}}.

\bibitem{Bottino:2008mf}
A.~Bottino, F.~Donato, N.~Fornengo, and S.~Scopel, ``{Interpreting the recent
  results on direct search for dark matter particles in terms of relic
  neutralino},'' \href{http://dx.doi.org/10.1103/PhysRevD.78.083520}{{\em Phys.
  Rev.} {\bfseries D78} (2008) 083520},
\href{http://arxiv.org/abs/0806.4099}{{\ttfamily arXiv:0806.4099 [hep-ph]}}.

\bibitem{Feldman:2010ke}
D.~Feldman, Z.~Liu, and P.~Nath, ``{Low Mass Neutralino Dark Matter in the MSSM
  with Constraints from $B_s\to \mu^+\mu^-$ and Higgs Search Limits},''
  \href{http://dx.doi.org/10.1103/PhysRevD.81.117701}{{\em Phys. Rev.}
  {\bfseries D81} (2010) 117701},
\href{http://arxiv.org/abs/1003.0437}{{\ttfamily arXiv:1003.0437 [hep-ph]}}.

\bibitem{Kuflik:2010ah}
E.~Kuflik, A.~Pierce, and K.~M. Zurek, ``{Light Neutralinos with Large
  Scattering Cross Sections in the Minimal Supersymmetric Standard Model},''
  \href{http://dx.doi.org/10.1103/PhysRevD.81.111701}{{\em Phys. Rev.}
  {\bfseries D81} (2010) 111701},
\href{http://arxiv.org/abs/1003.0682}{{\ttfamily arXiv:1003.0682 [hep-ph]}}.

\bibitem{Mandic:2000jz}
V.~Mandic, A.~Pierce, P.~Gondolo, and H.~Murayama, ``{The Lower bound on the
  neutralino nucleon cross-section},''
\href{http://arxiv.org/abs/hep-ph/0008022}{{\ttfamily arXiv:hep-ph/0008022}}.

\bibitem{Goodman:2010yf}
J.~Goodman {\em et al.}, ``{Constraints on Light Majorana Dark Matter from
  Colliders},'' \href{http://dx.doi.org/10.1016/j.physletb.2010.11.009}{{\em
  Phys. Lett.} {\bfseries B695} (2011) 185--188},
\href{http://arxiv.org/abs/1005.1286}{{\ttfamily arXiv:1005.1286 [hep-ph]}}.

\bibitem{Goodman:2010ku}
J.~Goodman {\em et al.}, ``{Constraints on Dark Matter from Colliders},''
  \href{http://dx.doi.org/10.1103/PhysRevD.82.116010}{{\em Phys. Rev.}
  {\bfseries D82} (2010) 116010},
\href{http://arxiv.org/abs/1008.1783}{{\ttfamily arXiv:1008.1783 [hep-ph]}}.

\bibitem{Bai:2010hh}
Y.~Bai, P.~J. Fox, and R.~Harnik, ``{The Tevatron at the Frontier of Dark
  Matter Direct Detection},''
  \href{http://dx.doi.org/10.1007/JHEP12(2010)048}{{\em JHEP} {\bfseries 12}
  (2010) 048},
\href{http://arxiv.org/abs/1005.3797}{{\ttfamily arXiv:1005.3797 [hep-ph]}}.

\bibitem{Fan:2010gt}
J.~Fan, M.~Reece, and L.-T. Wang, ``{Non-relativistic effective theory of dark
  matter direct detection},''
  \href{http://dx.doi.org/10.1088/1475-7516/2010/11/042}{{\em JCAP} {\bfseries
  1011} (2010) 042},
\href{http://arxiv.org/abs/1008.1591}{{\ttfamily arXiv:1008.1591 [hep-ph]}}.

\bibitem{Chamseddine:1982jx}
A.~H. Chamseddine, R.~L. Arnowitt, and P.~Nath, ``{Locally Supersymmetric Grand
  Unification},''
\href{http://dx.doi.org/10.1103/PhysRevLett.49.970}{{\em Phys. Rev. Lett.}
  {\bfseries 49} (1982) 970}.

\bibitem{Barbieri:1982eh}
R.~Barbieri, S.~Ferrara, and C.~A. Savoy, ``{Gauge Models with Spontaneously
  Broken Local Supersymmetry},''
\href{http://dx.doi.org/10.1016/0370-2693(82)90685-2}{{\em Phys. Lett.}
  {\bfseries B119} (1982) 343}.

\bibitem{Ohta:1982wn}
N.~Ohta, ``{Grand unified theories based on local supersymmetry},''
\href{http://dx.doi.org/10.1143/PTP.70.542}{{\em Prog. Theor. Phys.} {\bfseries
  70} (1983) 542}.

\bibitem{Hall:1983iz}
L.~J. Hall, J.~D. Lykken, and S.~Weinberg, ``{Supergravity as the Messenger of
  Supersymmetry Breaking},''
\href{http://dx.doi.org/10.1103/PhysRevD.27.2359}{{\em Phys. Rev.} {\bfseries
  D27} (1983) 2359--2378}.

\bibitem{AlvarezGaume:1983gj}
L.~Alvarez-Gaume, J.~Polchinski, and M.~B. Wise, ``{Minimal Low-Energy
  Supergravity},''
\href{http://dx.doi.org/10.1016/0550-3213(83)90591-6}{{\em Nucl. Phys.}
  {\bfseries B221} (1983) 495}.

\bibitem{Ellis:1999mm}
J.~R. Ellis, T.~Falk, K.~A. Olive, and M.~Srednicki, ``{Calculations of
  neutralino stau coannihilation channels and the cosmologically relevant
  region of MSSM parameter space},''
  \href{http://dx.doi.org/10.1016/S0927-6505(99)00104-8}{{\em Astropart. Phys.}
  {\bfseries 13} (2000) 181--213},
\href{http://arxiv.org/abs/hep-ph/9905481}{{\ttfamily arXiv:hep-ph/9905481}}.

\bibitem{Baer:2002fv}
H.~Baer, C.~Balazs, and A.~Belyaev, ``{Neutralino relic density in minimal
  supergravity with co- annihilations},'' {\em JHEP} {\bfseries 03} (2002) 042,
\href{http://arxiv.org/abs/hep-ph/0202076}{{\ttfamily arXiv:hep-ph/0202076}}.

\bibitem{Feng:2000gh}
J.~L. Feng, K.~T. Matchev, and F.~Wilczek, ``{Neutralino Dark Matter in Focus
  Point Supersymmetry},''
  \href{http://dx.doi.org/10.1016/S0370-2693(00)00512-8}{{\em Phys. Lett.}
  {\bfseries B482} (2000) 388--399},
\href{http://arxiv.org/abs/hep-ph/0004043}{{\ttfamily arXiv:hep-ph/0004043}}.

\bibitem{Feng:2000zu}
J.~L. Feng, K.~T. Matchev, and F.~Wilczek, ``{Prospects for indirect detection
  of neutralino dark matter},''
  \href{http://dx.doi.org/10.1103/PhysRevD.63.045024}{{\em Phys. Rev.}
  {\bfseries D63} (2001) 045024},
\href{http://arxiv.org/abs/astro-ph/0008115}{{\ttfamily
  arXiv:astro-ph/0008115}}.

\bibitem{Baer:2005ky}
H.~Baer, T.~Krupovnickas, S.~Profumo, and P.~Ullio, ``{Model independent
  approach to focus point supersymmetry: From dark matter to collider
  searches},'' \href{http://dx.doi.org/10.1088/1126-6708/2005/10/020}{{\em
  JHEP} {\bfseries 10} (2005) 020},
\href{http://arxiv.org/abs/hep-ph/0507282}{{\ttfamily arXiv:hep-ph/0507282}}.

\bibitem{Ellis:2001msa}
J.~R. Ellis, T.~Falk, G.~Ganis, K.~A. Olive, and M.~Srednicki, ``{The CMSSM
  Parameter Space at Large tan beta},''
  \href{http://dx.doi.org/10.1016/S0370-2693(01)00541-X}{{\em Phys. Lett.}
  {\bfseries B510} (2001) 236--246},
\href{http://arxiv.org/abs/hep-ph/0102098}{{\ttfamily arXiv:hep-ph/0102098}}.

\bibitem{deAustri:2006pe}
R.~R. de~Austri, R.~Trotta, and L.~Roszkowski, ``{A Markov chain Monte Carlo
  analysis of the CMSSM},'' {\em JHEP} {\bfseries 05} (2006) 002,
\href{http://arxiv.org/abs/hep-ph/0602028}{{\ttfamily arXiv:hep-ph/0602028}}.

\bibitem{Trotta:2006ew}
R.~Trotta, R.~R. de~Austri, and L.~Roszkowski, ``{Prospects for direct dark
  matter detection in the constrained MSSM},''
  \href{http://dx.doi.org/10.1016/j.newar.2006.11.059}{{\em New Astron. Rev.}
  {\bfseries 51} (2007) 316--320},
\href{http://arxiv.org/abs/astro-ph/0609126}{{\ttfamily
  arXiv:astro-ph/0609126}}.

\bibitem{Trotta:2008bp}
R.~Trotta, F.~Feroz, M.~P. Hobson, L.~Roszkowski, and R.~Ruiz~de Austri, ``{The
  Impact of priors and observables on parameter inferences in the Constrained
  MSSM},'' \href{http://dx.doi.org/10.1088/1126-6708/2008/12/024}{{\em JHEP}
  {\bfseries 12} (2008) 024},
\href{http://arxiv.org/abs/0809.3792}{{\ttfamily arXiv:0809.3792 [hep-ph]}}.

\bibitem{Baer:2006te}
H.~Baer, A.~Mustafayev, E.-K. Park, and X.~Tata, ``{Target dark matter
  detection rates in models with a well- tempered neutralino},'' {\em JCAP}
  {\bfseries 0701} (2007) 017,
\href{http://arxiv.org/abs/hep-ph/0611387}{{\ttfamily arXiv:hep-ph/0611387}}.

\bibitem{Djouadi:1998di}
{\bfseries MSSM Working Group} Collaboration, A.~Djouadi {\em et al.}, ``{The
  Minimal supersymmetric standard model: Group summary report},''
\href{http://arxiv.org/abs/hep-ph/9901246}{{\ttfamily arXiv:hep-ph/9901246}}.

\bibitem{Djouadi:2002ze}
A.~Djouadi, J.-L. Kneur, and G.~Moultaka, ``{SuSpect: A Fortran code for the
  supersymmetric and Higgs particle spectrum in the MSSM},''
  \href{http://dx.doi.org/10.1016/j.cpc.2006.11.009}{{\em Comput. Phys.
  Commun.} {\bfseries 176} (2007) 426--455},
\href{http://arxiv.org/abs/hep-ph/0211331}{{\ttfamily arXiv:hep-ph/0211331}}.

\bibitem{Berger:2008cq}
C.~F. Berger, J.~S. Gainer, J.~L. Hewett, and T.~G. Rizzo, ``{Supersymmetry
  Without Prejudice},''
  \href{http://dx.doi.org/10.1088/1126-6708/2009/02/023}{{\em JHEP} {\bfseries
  02} (2009) 023},
\href{http://arxiv.org/abs/0812.0980}{{\ttfamily arXiv:0812.0980 [hep-ph]}}.

\bibitem{Cotta:2009zu}
R.~C. Cotta, J.~S. Gainer, J.~L. Hewett, and T.~G. Rizzo, ``{Dark Matter in the
  MSSM},'' \href{http://dx.doi.org/10.1088/1367-2630/11/10/105026}{{\em New J.
  Phys.} {\bfseries 11} (2009) 105026},
\href{http://arxiv.org/abs/0903.4409}{{\ttfamily arXiv:0903.4409 [hep-ph]}}.

\bibitem{Allanach:2001kg}
B.~C. Allanach, ``{SOFTSUSY: A C++ program for calculating supersymmetric
  spectra},'' \href{http://dx.doi.org/10.1016/S0010-4655(01)00460-X}{{\em
  Comput. Phys. Commun.} {\bfseries 143} (2002) 305--331},
\href{http://arxiv.org/abs/hep-ph/0104145}{{\ttfamily arXiv:hep-ph/0104145}}.

\bibitem{Belanger:2006is}
G.~Belanger, F.~Boudjema, A.~Pukhov, and A.~Semenov, ``{micrOMEGAs2.0: A
  program to calculate the relic density of dark matter in a generic model},''
  \href{http://dx.doi.org/10.1016/j.cpc.2006.11.008}{{\em Comput. Phys.
  Commun.} {\bfseries 176} (2007) 367--382},
\href{http://arxiv.org/abs/hep-ph/0607059}{{\ttfamily arXiv:hep-ph/0607059}}.

\bibitem{Belanger:2008sj}
G.~Belanger, F.~Boudjema, A.~Pukhov, and A.~Semenov, ``{Dark matter direct
  detection rate in a generic model with micrOMEGAs2.1},''
  \href{http://dx.doi.org/10.1016/j.cpc.2008.11.019}{{\em Comput. Phys.
  Commun.} {\bfseries 180} (2009) 747--767},
\href{http://arxiv.org/abs/0803.2360}{{\ttfamily arXiv:0803.2360 [hep-ph]}}.

\bibitem{Mizuta:1992ja}
S.~Mizuta, D.~Ng, and M.~Yamaguchi, ``{Phenomenological aspects of
  supersymmetric standard models without grand unification},''
  \href{http://dx.doi.org/10.1016/0370-2693(93)90754-6}{{\em Phys. Lett.}
  {\bfseries B300} (1993) 96--103},
\href{http://arxiv.org/abs/hep-ph/9210241}{{\ttfamily arXiv:hep-ph/9210241}}.

\bibitem{Baer:2006dz}
H.~Baer, A.~Mustafayev, E.-K. Park, S.~Profumo, and X.~Tata, ``{Mixed higgsino
  dark matter from a reduced SU(3) gaugino mass: Consequences for dark matter
  and collider searches},''
  \href{http://dx.doi.org/10.1088/1126-6708/2006/04/041}{{\em JHEP} {\bfseries
  04} (2006) 041},
\href{http://arxiv.org/abs/hep-ph/0603197}{{\ttfamily arXiv:hep-ph/0603197}}.

\bibitem{Baer:2007xd}
H.~Baer, A.~Mustafayev, H.~Summy, and X.~Tata, ``{Mixed Higgsino Dark Matter
  from a Large SU(2) Gaugino Mass},''
  \href{http://dx.doi.org/10.1088/1126-6708/2007/10/088}{{\em JHEP} {\bfseries
  10} (2007) 088},
\href{http://arxiv.org/abs/0708.4003}{{\ttfamily arXiv:0708.4003 [hep-ph]}}.

\bibitem{Anderson:1996bg}
G.~Anderson {\em et al.}, ``{Motivations for and implications of non-universal
  GUT- scale boundary conditions for soft SUSY-breaking parameters},''
\href{http://arxiv.org/abs/hep-ph/9609457}{{\ttfamily arXiv:hep-ph/9609457}}.

\bibitem{Chattopadhyay:2001va}
U.~Chattopadhyay, A.~Corsetti, and P.~Nath, ``{Supersymmetric dark matter and
  Yukawa unification},''
  \href{http://dx.doi.org/10.1103/PhysRevD.66.035003}{{\em Phys. Rev.}
  {\bfseries D66} (2002) 035003},
\href{http://arxiv.org/abs/hep-ph/0201001}{{\ttfamily arXiv:hep-ph/0201001}}.

\bibitem{Ellis:2002wv}
J.~R. Ellis, K.~A. Olive, and Y.~Santoso, ``{The MSSM Parameter Space with
  Non-Universal Higgs Masses},''
  \href{http://dx.doi.org/10.1016/S0370-2693(02)02071-3}{{\em Phys. Lett.}
  {\bfseries B539} (2002) 107--118},
\href{http://arxiv.org/abs/hep-ph/0204192}{{\ttfamily arXiv:hep-ph/0204192}}.

\bibitem{BirkedalHansen:2002am}
A.~Birkedal-Hansen and B.~D. Nelson, ``{Relic neutralino densities and
  detection rates with nonuniversal gaugino masses},''
  \href{http://dx.doi.org/10.1103/PhysRevD.67.095006}{{\em Phys. Rev.}
  {\bfseries D67} (2003) 095006},
\href{http://arxiv.org/abs/hep-ph/0211071}{{\ttfamily arXiv:hep-ph/0211071}}.

\bibitem{Ellis:2003eg}
J.~R. Ellis, A.~Ferstl, K.~A. Olive, and Y.~Santoso, ``{Direct Detection of
  Dark Matter in the MSSM with Non- Universal Higgs Masses},''
  \href{http://dx.doi.org/10.1103/PhysRevD.67.123502}{{\em Phys. Rev.}
  {\bfseries D67} (2003) 123502},
\href{http://arxiv.org/abs/hep-ph/0302032}{{\ttfamily arXiv:hep-ph/0302032}}.

\bibitem{Baer:2005bu}
H.~Baer, A.~Mustafayev, S.~Profumo, A.~Belyaev, and X.~Tata, ``{Direct,
  indirect and collider detection of neutralino dark matter in SUSY models with
  non-universal Higgs masses},'' {\em JHEP} {\bfseries 07} (2005) 065,
\href{http://arxiv.org/abs/hep-ph/0504001}{{\ttfamily arXiv:hep-ph/0504001}}.

\bibitem{Feng:2001sq}
J.~L. Feng, K.~T. Matchev, and Y.~Shadmi, ``{Theoretical Expectations for the
  Muon's Electric Dipole Moment},''
  \href{http://dx.doi.org/10.1016/S0550-3213(01)00383-2}{{\em Nucl. Phys.}
  {\bfseries B613} (2001) 366--381},
\href{http://arxiv.org/abs/hep-ph/0107182}{{\ttfamily arXiv:hep-ph/0107182}}.

\bibitem{Moroi:1995yh}
T.~Moroi, ``{The Muon Anomalous Magnetic Dipole Moment in the Minimal
  Supersymmetric Standard Model},''
  \href{http://dx.doi.org/10.1103/PhysRevD.53.6565}{{\em Phys. Rev.} {\bfseries
  D53} (1996) 6565--6575},
\href{http://arxiv.org/abs/hep-ph/9512396}{{\ttfamily arXiv:hep-ph/9512396}}.

\bibitem{Regan:2002ta}
B.~C. Regan, E.~D. Commins, C.~J. Schmidt, and D.~DeMille, ``{New limit on the
  electron electric dipole moment},''
\href{http://dx.doi.org/10.1103/PhysRevLett.88.071805}{{\em Phys. Rev. Lett.}
  {\bfseries 88} (2002) 071805}.

\bibitem{Baker:2006ts}
C.~A. Baker {\em et al.}, ``{An improved experimental limit on the electric
  dipole moment of the neutron},''
  \href{http://dx.doi.org/10.1103/PhysRevLett.97.131801}{{\em Phys. Rev. Lett.}
  {\bfseries 97} (2006) 131801},
\href{http://arxiv.org/abs/hep-ex/0602020}{{\ttfamily arXiv:hep-ex/0602020}}.

\bibitem{Bennett:2006fi}
{\bfseries Muon G-2} Collaboration, G.~W. Bennett {\em et al.}, ``{Final report
  of the muon E821 anomalous magnetic moment measurement at BNL},''
  \href{http://dx.doi.org/10.1103/PhysRevD.73.072003}{{\em Phys. Rev.}
  {\bfseries D73} (2006) 072003},
\href{http://arxiv.org/abs/hep-ex/0602035}{{\ttfamily arXiv:hep-ex/0602035}}.

\bibitem{Barberio:2007cr}
{\bfseries Heavy Flavor Averaging Group (HFAG)} Collaboration, E.~Barberio {\em
  et al.}, ``{Averages of $b-$hadron properties at the end of 2006},''
\href{http://arxiv.org/abs/0704.3575}{{\ttfamily arXiv:0704.3575 [hep-ex]}}.

\bibitem{Aaltonen:2007kv}
{\bfseries CDF} Collaboration, T.~Aaltonen {\em et al.}, ``{Search for $B^0_{s}
  \to \mu^{+} \mu^{-}$ and $B^0_{d} \to \mu^{+} \mu^{-}$ decays with $2fb^{-1}$
  of $p \bar{p}$ collisions},''
  \href{http://dx.doi.org/10.1103/PhysRevLett.100.101802}{{\em Phys. Rev.
  Lett.} {\bfseries 100} (2008) 101802},
\href{http://arxiv.org/abs/0712.1708}{{\ttfamily arXiv:0712.1708 [hep-ex]}}.

\bibitem{Misiak:2006zs}
M.~Misiak {\em et al.}, ``{The first estimate of $B(\bar{B} \to X_s \gamma)$ at
  ${\cal O}(\alpha_s^2)$},''
  \href{http://dx.doi.org/10.1103/PhysRevLett.98.022002}{{\em Phys. Rev. Lett.}
  {\bfseries 98} (2007) 022002},
\href{http://arxiv.org/abs/hep-ph/0609232}{{\ttfamily arXiv:hep-ph/0609232}}.

\bibitem{Misiak:2006ab}
M.~Misiak and M.~Steinhauser, ``{NNLO QCD corrections to the $B \to X_s \gamma$
  matrix elements using interpolation in $m_c$},''
  \href{http://dx.doi.org/10.1016/j.nuclphysb.2006.11.027}{{\em Nucl. Phys.}
  {\bfseries B764} (2007) 62--82},
\href{http://arxiv.org/abs/hep-ph/0609241}{{\ttfamily arXiv:hep-ph/0609241}}.

\bibitem{Jegerlehner:2009ry}
F.~Jegerlehner and A.~Nyffeler, ``{The Muon g-2},''
  \href{http://dx.doi.org/10.1016/j.physrep.2009.04.003}{{\em Phys. Rept.}
  {\bfseries 477} (2009) 1--110},
\href{http://arxiv.org/abs/0902.3360}{{\ttfamily arXiv:0902.3360 [hep-ph]}}.

\bibitem{Dine:2007xi}
M.~Dine, N.~Seiberg, and S.~Thomas, ``{Higgs Physics as a Window Beyond the
  MSSM (BMSSM)},'' \href{http://dx.doi.org/10.1103/PhysRevD.76.095004}{{\em
  Phys. Rev.} {\bfseries D76} (2007) 095004},
\href{http://arxiv.org/abs/0707.0005}{{\ttfamily arXiv:0707.0005 [hep-ph]}}.

\bibitem{Falk:1995cq}
T.~Falk, K.~A. Olive, L.~Roszkowski, and M.~Srednicki, ``{New Constraints on
  Superpartner Masses},''
  \href{http://dx.doi.org/10.1016/0370-2693(95)01430-6}{{\em Phys. Lett.}
  {\bfseries B367} (1996) 183--187},
\href{http://arxiv.org/abs/hep-ph/9510308}{{\ttfamily arXiv:hep-ph/9510308}}.

\bibitem{Kusenko:1996jn}
A.~Kusenko, P.~Langacker, and G.~Segre, ``{Phase Transitions and Vacuum
  Tunneling Into Charge and Color Breaking Minima in the MSSM},''
  \href{http://dx.doi.org/10.1103/PhysRevD.54.5824}{{\em Phys. Rev.} {\bfseries
  D54} (1996) 5824--5834},
\href{http://arxiv.org/abs/hep-ph/9602414}{{\ttfamily arXiv:hep-ph/9602414}}.

\bibitem{Feng:2005ba}
J.~L. Feng, A.~Rajaraman, and B.~T. Smith, ``{Minimal supergravity with $m_0^2
  < 0$},'' \href{http://dx.doi.org/10.1103/PhysRevD.74.015013}{{\em Phys. Rev.}
  {\bfseries D74} (2006) 015013},
\href{http://arxiv.org/abs/hep-ph/0512172}{{\ttfamily arXiv:hep-ph/0512172}}.

\bibitem{Rajaraman:2006mr}
A.~Rajaraman and B.~T. Smith, ``{Discovering SUSY with $m_0^2 < 0$ in the first
  CERN LHC physics run},''
  \href{http://dx.doi.org/10.1103/PhysRevD.75.115015}{{\em Phys. Rev.}
  {\bfseries D75} (2007) 115015},
\href{http://arxiv.org/abs/hep-ph/0612235}{{\ttfamily arXiv:hep-ph/0612235}}.

\bibitem{Ellis:2008mc}
J.~R. Ellis, J.~Giedt, O.~Lebedev, K.~Olive, and M.~Srednicki, ``{Against
  Tachyophobia},'' \href{http://dx.doi.org/10.1103/PhysRevD.78.075006}{{\em
  Phys. Rev.} {\bfseries D78} (2008) 075006},
\href{http://arxiv.org/abs/0806.3648}{{\ttfamily arXiv:0806.3648 [hep-ph]}}.

\bibitem{Evans:2008zx}
J.~L. Evans, D.~E. Morrissey, and J.~D. Wells, ``{Vacuum Stability with
  Tachyonic Boundary Higgs Masses in No-Scale Supersymmetry or Gaugino
  Mediation},'' \href{http://dx.doi.org/10.1103/PhysRevD.80.095011}{{\em Phys.
  Rev.} {\bfseries D80} (2009) 095011},
\href{http://arxiv.org/abs/0812.3874}{{\ttfamily arXiv:0812.3874 [hep-ph]}}.

\bibitem{ElKheishen:1992yv}
M.~M. El~Kheishen, A.~A. Aboshousha, and A.~A. Shafik, ``{Analytic formulas for
  the neutralino masses and the neutralino mixing matrix},''
\href{http://dx.doi.org/10.1103/PhysRevD.45.4345}{{\em Phys. Rev.} {\bfseries
  D45} (1992) 4345--4348}.

\bibitem{ArkaniHamed:2006mb}
N.~Arkani-Hamed, A.~Delgado, and G.~F. Giudice, ``{The well-tempered
  neutralino},'' \href{http://dx.doi.org/10.1016/j.nuclphysb.2006.02.010}{{\em
  Nucl. Phys.} {\bfseries B741} (2006) 108--130},
\href{http://arxiv.org/abs/hep-ph/0601041}{{\ttfamily arXiv:hep-ph/0601041}}.

\bibitem{Shifman:1978zn}
M.~A. Shifman, A.~I. Vainshtein, and V.~I. Zakharov, ``{Remarks on Higgs Boson
  Interactions with Nucleons},''
\href{http://dx.doi.org/10.1016/0370-2693(78)90481-1}{{\em Phys. Lett.}
  {\bfseries B78} (1978) 443}.

\bibitem{Giedt:2009mr}
J.~Giedt, A.~W. Thomas, and R.~D. Young, ``{Dark matter, the CMSSM and lattice
  QCD},'' \href{http://dx.doi.org/10.1103/PhysRevLett.103.201802}{{\em Phys.
  Rev. Lett.} {\bfseries 103} (2009) 201802},
\href{http://arxiv.org/abs/0907.4177}{{\ttfamily arXiv:0907.4177 [hep-ph]}}.

\bibitem{Ellis:2000jd}
J.~R. Ellis, A.~Ferstl, and K.~A. Olive, ``{Exploration of elastic scattering
  rates for supersymmetric dark matter},''
  \href{http://dx.doi.org/10.1103/PhysRevD.63.065016}{{\em Phys. Rev.}
  {\bfseries D63} (2001) 065016},
\href{http://arxiv.org/abs/hep-ph/0007113}{{\ttfamily arXiv:hep-ph/0007113}}.

\bibitem{Falk:1999mq}
T.~Falk, A.~Ferstl, and K.~A. Olive, ``{Variations of the neutralino elastic
  cross-section with CP violating phases},''
  \href{http://dx.doi.org/10.1016/S0927-6505(99)00125-5}{{\em Astropart. Phys.}
  {\bfseries 13} (2000) 301--316},
\href{http://arxiv.org/abs/hep-ph/9908311}{{\ttfamily arXiv:hep-ph/9908311}}.

\bibitem{Feng:2009te}
J.~L. Feng, J.-F. Grivaz, and J.~Nachtman, ``{Searches for Supersymmetry at
  High-Energy Colliders},''
  \href{http://dx.doi.org/10.1103/RevModPhys.82.699}{{\em Rev. Mod. Phys.}
  {\bfseries 82} (2010) 699--727},
\href{http://arxiv.org/abs/0903.0046}{{\ttfamily arXiv:0903.0046 [hep-ex]}}.

\bibitem{Fiorucci:2009ak}
S.~Fiorucci {\em et al.}, ``{Status of the LUX Dark Matter Search},''
  \href{http://dx.doi.org/10.1063/1.3327777}{{\em AIP Conf. Proc.} {\bfseries
  1200} (2010) 977--980},
\href{http://arxiv.org/abs/0912.0482}{{\ttfamily arXiv:0912.0482
  [astro-ph.CO]}}.

\bibitem{LUXwebsite}
``{LUX Dark Matter Experiment website}.''
\newblock \url{http://lux.brown.edu}.

\bibitem{Dimopoulos:1995mi}
S.~Dimopoulos and G.~F. Giudice, ``{Naturalness constraints in supersymmetric
  theories with nonuniversal soft terms},''
  \href{http://dx.doi.org/10.1016/0370-2693(95)00961-J}{{\em Phys. Lett.}
  {\bfseries B357} (1995) 573--578},
\href{http://arxiv.org/abs/hep-ph/9507282}{{\ttfamily arXiv:hep-ph/9507282}}.

\bibitem{Dvali:1996rj}
G.~R. Dvali and A.~Pomarol, ``{Anomalous U(1) as a mediator of supersymmetry
  breaking},'' \href{http://dx.doi.org/10.1103/PhysRevLett.77.3728}{{\em Phys.
  Rev. Lett.} {\bfseries 77} (1996) 3728--3731},
\href{http://arxiv.org/abs/hep-ph/9607383}{{\ttfamily arXiv:hep-ph/9607383}}.

\bibitem{Cohen:1996vb}
A.~G. Cohen, D.~B. Kaplan, and A.~E. Nelson, ``{The more minimal supersymmetric
  standard model},''
  \href{http://dx.doi.org/10.1016/S0370-2693(96)01183-5}{{\em Phys. Lett.}
  {\bfseries B388} (1996) 588--598},
\href{http://arxiv.org/abs/hep-ph/9607394}{{\ttfamily arXiv:hep-ph/9607394}}.

\bibitem{Feng:1998iq}
J.~L. Feng, C.~F. Kolda, and N.~Polonsky, ``{Solving the supersymmetric flavor
  problem with radiatively generated mass hierarchies},''
  \href{http://dx.doi.org/10.1016/S0550-3213(99)00026-7}{{\em Nucl. Phys.}
  {\bfseries B546} (1999) 3--18},
\href{http://arxiv.org/abs/hep-ph/9810500}{{\ttfamily arXiv:hep-ph/9810500}}.

\bibitem{Bagger:1999ty}
J.~Bagger, J.~L. Feng, and N.~Polonsky, ``{Naturally heavy scalars in
  supersymmetric grand unified theories},''
  \href{http://dx.doi.org/10.1016/S0550-3213(99)00577-5}{{\em Nucl. Phys.}
  {\bfseries B563} (1999) 3--20},
\href{http://arxiv.org/abs/hep-ph/9905292}{{\ttfamily arXiv:hep-ph/9905292}}.

\bibitem{Nihei:2004bc}
T.~Nihei and M.~Sasagawa, ``{Relic density and elastic scattering cross
  sections of the neutralino in the MSSM with CP-violating phases},''
  \href{http://dx.doi.org/10.1103/PhysRevD.70.055011}{{\em Phys. Rev.}
  {\bfseries D70} (2004) 055011},
\href{http://arxiv.org/abs/hep-ph/0404100}{{\ttfamily arXiv:hep-ph/0404100}}.

\bibitem{Belanger:2006qa}
G.~Belanger, F.~Boudjema, S.~Kraml, A.~Pukhov, and A.~Semenov, ``{Relic density
  of neutralino dark matter in the MSSM with CP violation},''
  \href{http://dx.doi.org/10.1103/PhysRevD.73.115007}{{\em Phys. Rev.}
  {\bfseries D73} (2006) 115007},
\href{http://arxiv.org/abs/hep-ph/0604150}{{\ttfamily arXiv:hep-ph/0604150}}.

\bibitem{Kamionkowski:2008vw}
M.~Kamionkowski and S.~M. Koushiappas, ``{Galactic Substructure and Direct
  Detection of Dark Matter},''
  \href{http://dx.doi.org/10.1103/PhysRevD.77.103509}{{\em Phys. Rev.}
  {\bfseries D77} (2008) 103509},
\href{http://arxiv.org/abs/0801.3269}{{\ttfamily arXiv:0801.3269 [astro-ph]}}.

\bibitem{Achterberg:2006jf}
{\bfseries AMANDA} Collaboration, A.~Achterberg {\em et al.}, ``{Limits on the
  muon flux from neutralino annihilations at the center of the Earth with
  AMANDA},''
\href{http://dx.doi.org/10.1016/j.astropartphys.2006.05.007}{{\em Astropart.
  Phys.} {\bfseries 26} (2006) 129--139}.

\bibitem{Abbasi:2009uz}
{\bfseries IceCube} Collaboration, R.~Abbasi {\em et al.}, ``{Limits on a muon
  flux from neutralino annihilations in the Sun with the IceCube 22-string
  detector},'' \href{http://dx.doi.org/10.1103/PhysRevLett.102.201302}{{\em
  Phys. Rev. Lett.} {\bfseries 102} (2009) 201302},
\href{http://arxiv.org/abs/0902.2460}{{\ttfamily arXiv:0902.2460
  [astro-ph.CO]}}.

\bibitem{Braun:2009fr}
{\bfseries IceCube} Collaboration, J.~Braun, D.~Hubert, {\em et al.},
  ``{Searches for WIMP Dark Matter from the Sun with AMANDA},''
\href{http://arxiv.org/abs/0906.1615}{{\ttfamily arXiv:0906.1615
  [astro-ph.HE]}}.

\bibitem{Behnke:2010xt}
E.~Behnke {\em et al.}, ``{Improved Limits on Spin-Dependent WIMP-Proton
  Interactions from a Two Liter CF$_3$I Bubble Chamber},''
  \href{http://dx.doi.org/10.1103/PhysRevLett.106.021303}{{\em Phys. Rev.
  Lett.} {\bfseries 106} (2011) 021303},
\href{http://arxiv.org/abs/1008.3518}{{\ttfamily arXiv:1008.3518
  [astro-ph.CO]}}.

\end{thebibliography}\endgroup

\end{document}